\documentclass[a4paper,prd,11pt,noshowkeys,notitlepage,nofootinbib,superscriptaddress]{revtex4}
\usepackage[utf8]{inputenc}
\usepackage[T1]{fontenc}
\usepackage{ae,aecompl}
\usepackage{amsmath,amssymb,mathrsfs}
\usepackage{graphicx}
\usepackage{dsfont}
\usepackage{slashed}
\usepackage{units}
\usepackage[version=4]{mhchem}
\usepackage[normalem]{ulem}
\usepackage[usenames,dvipsnames]{xcolor} 
\usepackage{hyperref}
\hypersetup{
    colorlinks=true,        
    linkcolor=purple,       
    citecolor=JungleGreen,  
    filecolor=magenta,      
}
\usepackage{xfrac}
\usepackage{ccnishi}
\providecommand{\lag}{\mathscr{L}} 
\providecommand{\Br}{\operatorname{Br}} 

\providecommand{\tPhi}{\tilde{\Phi}}
\providecommand{\tM}{\tilde{M}}

\providecommand{\la}[1]{\lambda^{(#1)}}

\providecommand{\La}[1]{\Lambda^{(#1)}}
\providecommand{\LGN}{\Lambda_{\rm GN}}
\providecommand{\muphi}{\mu_\varphi}
\newcommand{\amu}{\ensuremath{a_\mu}}
 
\providecommand{\eps}{\epsilon} 

\providecommand{\cH}{\mathcal{H}}
\providecommand{\tf}{\tilde{f}}

\providecommand{\GFSM}{G_F^{\rm SM}}
\providecommand{\Azee}{A_{\rm Zee}}

\def\beq{\begin{equation}}
\def\eeq{\end{equation}}
\def\beqa{\begin{eqnarray}}
\def\eeqa{\end{eqnarray}}

\def\ifmath#1{\relax\ifmmode #1\else $#1$\fi}

\def\cbma{c_{\beta-\alpha}}

\def\sbma{s_{\beta-\alpha}}

\def\mhl{m_{\hl}}
\def\mhh{m_{\hh}}

\def\hl{h}

\def\hh{H}

\makeatletter
\DeclareRobustCommand{\cev}[1]{%
  {\mathpalette\do@cev{#1}}%
}
\newcommand{\do@cev}[2]{%
  \vbox{\offinterlineskip
    \sbox\z@{$\m@th#1 x$}%
    \ialign{##\cr
      \hidewidth\reflectbox{$\m@th#1\vec{}\mkern4mu$}\hidewidth\cr
      \noalign{\kern-\ht\z@}
      $\m@th#1#2$\cr
    }%
  }%
}
\makeatother

\makeatletter
\newcommand{\overleftrightsmallarrow}{\mathpalette{\overarrowsmall@\leftrightarrowfill@}}
\newcommand{\overrightsmallarrow}{\mathpalette{\overarrowsmall@\rightarrowfill@}}
\newcommand{\overleftsmallarrow}{\mathpalette{\overarrowsmall@\leftarrowfill@}}
\newcommand{\overarrowsmall@}[3]{%
  \vbox{%
    \ialign{%
      ##\crcr
      #1{\smaller@style{#2}}\crcr
      \noalign{\nointerlineskip}%
      $\m@th\hfil#2#3\hfil$\crcr
    }%
  }%
}
\def\smaller@style#1{%
  \ifx#1\displaystyle\scriptstyle\else
    \ifx#1\textstyle\scriptstyle\else
      \scriptscriptstyle
    \fi
  \fi
}
\makeatother
\newcommand{\oLR}[1]{\overleftrightsmallarrow{#1}} 
\begin{document}
\title{
Aligned Zee-Grimus-Neufeld model for $(g-2)_\mu$
}
\author{R.~E.~A.~Bringas}
\email{ruben.bringas@ufabc.edu.br}
\affiliation{Centro de Ci\^{e}ncias Naturais e Humanas,
Universidade Federal do ABC, Santo Andr\'{e}-SP, Brasil}
\email{alche@unicamp.br}
\author{A.~L.~Cherchiglia}
\email{alche@unicamp.br}
\affiliation{Instituto de F\'isica Gleb Wataghin, Universidade Estadual de Campinas, Campinas-SP, Brazil}
\author{G.~De~Conto}
\email{george.de.conto@gmail.com}
\affiliation{Centro de Ci\^{e}ncias Naturais e Humanas,
Universidade Federal do ABC, Santo Andr\'{e}-SP, Brasil}
\author{C.~C.~Nishi}
\email{celso.nishi@ufabc.edu.br}
\affiliation{Centro de Matem\'{a}tica, Computa\c{c}\~{a}o e Cogni\c{c}\~{a}o,
Universidade Federal do ABC, Santo Andr\'{e}-SP, Brasil}

\begin{abstract}
The recent result of the Fermilab Muon g-2 experiment is still in conflict with the SM prediction if dispersive methods 
are used as input for the hadronic vacuum polarization.
We seek models where the same set of mediators around the weak scale or above contribute to neutrino masses and to the 
muon $g-2$ through a chiral enhanced contribution.
As family lepton number violation is inherent to the former, non-negligible charged lepton flavor violation will be 
induced.
We arrive at the following field content additional to the SM: one Higgs doublet, one singly 
charged scalar and \emph{one} righthanded neutrino.
Neutrino masses will be generated by a combination of mechanisms: tree-level seesaw, Zee mechanism and Grimus-Neufeld 
mechanism while lepton flavor violation will be minimized by considering Yukawa aligned Higgs doublets.
The setting is flexible enough to account for $(g-2)_\mu$, the Cabibbo angle anomaly and the $W$ mass deviation
while avoiding dangerous CLFV.
\end{abstract}
\maketitle
\section{Introduction}

The observation of neutrino oscillations firmly established that neutrinos have tiny masses and mix in the weak charged 
current\,\cite{ohlsson}. These properties clearly demonstrate that family lepton numbers associated to the lepton 
flavors $e,\mu,\tau$ are not conserved in nature.
However, if both family lepton number violation and neutrino masses have their origin in a high scale physics
such as a seesaw scale of $10^{12}\,\unit{GeV}$, 
its effects leading to deviations of the SM are expected to be unobservably small.
These effects include those of charged lepton flavor violation (CLFV) that are expected to be probed with much greater 
precision in the coming years\,\cite{clfv:status}.

In contrast, radiative neutrino mass models naturally require the responsible new physics to be realized at lower 
scales than 
the usual tree-level completions of the seesaw mechanism.\footnote{%
See e.g.\,\cite{Zee:1980ai,Cheng:1980qt,Zee:1985id,Babu:1988ki,ma:seesaws,rabi.pal}.
}
These models naturally lead to a host of effects at lower energy\,\cite{cai}, including CLFV effects which are also 
brought to low scale.
In that context, the simplest possibilities to generate light neutrino masses at one-loop without additional 
symmetries 
consist in enlarging the SM with a second Higgs doublet and additionally with either (i) one singly charged 
scalar in the Zee mechanism \cite{Zee:1980ai} or, as a less known option, with (ii) \emph{one} righthanded neutrino 
(RHN) 
$N_R$ in the Grimus-Neufeld mechanism \cite{grimus.neufeld}.
The simplest possibilities considering dark symmetries lead to the well-known scotogenic model\,\cite{scoto} and 
variants such as the scoto-seesasaw model\,\cite{scoto-ss}.
See Ref.\,\cite{cai} for a survey of other models.

Among the experimental anomalies of the SM\,\cite{anomalies:sm}, the discrepancy between the experimental and SM 
prediction for the muon magnetic moment, $(g-2)_\mu$ or 
$\amu$, is a long-standing one. The experiment E989 conducted at Fermilab has recently released their final 
result\;\cite{Muong-2:2025xyk}, which has decreased the error in the past world average\;\cite{Muong-2:2006rrc} 
by a factor of four.
The combined result from both collaborations is given by
$
\amu^\text{exp} = (116 592 07.15\pm1.45)\times10^{-10}.
$
The SM prediction relies on two different approaches in order to deal with the hadronic vacuum polarization 
contribution (HVP). Using the results from lattice, the SM prediction is given by\;\cite{Aliberti:2025beg}
$
\amu^{\text{SM}}= (116 592 03.3\pm6.2)\times10^{-10}.
$ 
In contrast, the SM prediction using the results from dispersive methods is given 
by\;\cite{Aoyama:2020ynm,g-2:theory}\footnote{The recent result from the CMD-3 experiment\;\cite{CMD3} is in tension 
with previous measurements, which do not allow for a meaningful combination. Until the situation is settled, we opt to 
omit the CMD-3 experiment from the combination.} 
$
	\amu^{\text{SM}} = (11659181.0\pm4.3)\times10^{-10}.
$
By comparison of the SM prediction using dispersive methods (which is more precise\footnote{We also remark that the 
present lattice value for the HVP contribution leads to a set of tensions in the EW fit performed by the authors of 
Ref.\;\cite{crivellin:hvp}. See \cite{Aliberti:2025beg} for further discussions.}), one obtains the following 
discrepancy:
\begin{equation}
\delta\amu^{\text{BSM}} = (26.2\pm4.5)\times10^{-10}.
\label{eq:amuBSM}
\end{equation}
In this work, we will adopt the above value for $\delta\amu^{\text{BSM}}$, 
seeking minimal models that can explain such 
discrepancy with the same mediators that generate
neutrino masses, and yet avoiding stringent constraints from CLFV processes.

Typically a large contribution to $\delta\amu^{\text{BSM}}$ at the order of \eqref{eq:amuBSM} requires TeV scale 
(or lower) new physics coupling to the muon that may or may not violate lepton flavor.
For this reason, many new physics scenarios addressing it could be probed at muon colliders\,\cite{curtin,g-2:collider}.
By minimally introducing new physics that couples to the muon but is unrelated to neutrino masses in a flavor 
conserving fashion or obeying minimal flavor violation, one can explain 
$\amu$\,\cite{strumia,freitas,crivellin:g-2,calibbi.ziegler} and still easily avoid large CLFV effects.
A critical review of the minimal extensions with one or two new fields can be seen in Refs.\,\cite{stockinger,queiroz}.
If the dominant new physics enters above the electroweak scale and contributes at 1-loop, a key 
ingredient is the presence of a \emph{chiral 
enhancement} that occurs by replacing one chiral flipping coupling involving the muon Yukawa $y_\mu\sim 0.0006$ by an 
order one coupling\,\cite{crivellin:g-2,calibbi.ziegler,curtin,strumia}.

In the Two-Higgs-doublet model (2HDM), a major enhancement is through Barr-Zee 
contributions at two-loops involving light neutral higgses and third generation fermions.
This contribution helps explaining \eqref{eq:amuBSM} in the type Y/X 2HDM\,\cite{2hdm.typeX}, aligned 
2HDM\,\cite{Ilisie,Han,Cherchiglia1,Cherchiglia2} or the Zee 
model\,\cite{barman}. For the latter, a chiral enhancement through the tau Yukawa is also present at one-loop.
The exchange of RHNs plays no role.

Here we seek a different mechanism, similar to Ref.\,\cite{nu.g-2},
where the one-loop exchange of $N_R$ and charged higgses are crucial.
This chiral enhanced contribution allows the new fields to be above the weak scale at TeV.
As lepton flavor violation necessary for neutrino masses is built in with the TeV scale new physics, the induced CLFV 
is also expected to be large and its study will be one of our major goals.
Our field content additional to the SM is minimal: one Higgs doublet, one singly charged scalar and \emph{one} RHN.
The latter differs from Ref.\,\cite{nu.g-2} where two RHNs were considered.
Here also the Zee contribution to neutrino mass is active by switching on the couplings between the lepton doublets and 
the singly charged scalar, and the two Higgs doublets are Yukawa aligned to minimize lepton flavor violation.
Therefore we consider the possibility of many simultaneous sources for neutrino masses: the tree-level seesaw, the Zee 
mechanism and the Grimus-Neufeld mechanism.
Turning on the coupling between the singly charged singlet with the lepton doublets also introduces many 
interesting effects\,\cite{crivellin:singly.charged,herrero:singly.charged}: due to 
its antisymmetric nature, it is inherently lepton flavor violating and it can solve the Cabibbo angle anomaly.
We should also mention the difference with respect to the model in \cite{dicus.he} which contains one more charged 
singlet and the chiral enhanced contribution to $a_\mu$ is absent.
Another earlier attempt to connect neutrino masses with $a_\mu$ can be found in Ref.\,\cite{ma.raidal} but there was 
no chiral enhancement and the contribution had the wrong sign.
Other recent works connecting $a_\mu$ with neutrino mass generation can be seen in 
Ref.\,\cite{nu.g-2:others,Dcruz:2022dao,lq.neutrino}.

The outline of this paper is as follows:
in Sec.\,\ref{sec:models} we present the generic model of Yukawa-aligned 2HDM with the addition of one charged singlet 
scalar and a \emph{single} righthanded neutrino. 
Section \ref{sec:nu.mass} reviews the various mechanisms available to generate neutrino masses while 
Sec.\,\ref{sec:combining} analyzes the combinations of more than one mechanism.
The various constraints coming from flavor and precision electroweak observables are discussed in 
Sec.\,\ref{sec:contraints}.
The surviving model is analyzed in Sec.\,\ref{sec:results} with respect to $(g-2)_\mu$, CLFV, Cabibbo angle anomaly 
and $W$ mass deviation.
The summary is finally presented in Sec.\,\ref{sec:summary}.

\section{Models with Yukawa-aligned Higgs}
\label{sec:models}

In the Two-Higgs-Doublet model (2HDM), generic unsupressed flavor changing neutral currents (FCNC) mediated by scalars are expected already at tree level unless some symmetry such as $\ZZ_2$ or other protection is enforced.
The flavor alignment of Yukawa couplings\,\cite{pich.tuzon} is one simple way of suppressing FCNC which is radiatively stable (within the 2HDM).
In this setting, the quark Yukawa couplings $\Gamma_i,\Delta_i$ in
\eq{
-\lag=\bar{q}_L(\Gamma_1\Phi_1+\Gamma_2\Phi_2)d_R+\bar{q}_L(\Delta_1\tPhi_1+\Delta_2\tPhi_2)u_R\,,
}
are aligned in flavor space as $\Gamma_2\propto \Gamma_1$ and $\Delta_2\propto \Delta_1$.
This alignment ensures that the Yukawa coupling responsible for quark masses can be simultaneously diagonalized with the second Yukawa coupling controlling the Yukawa interactions of the extra scalars.

Here we consider the Yukawa-aligned 2HDM, with two Higgs doublets $\Phi_1, \Phi_2$, 
adjoined with \emph{one}
righthanded neutrino (RHN) $N_{R}$ and \emph{one} singlet charged scalar $\varphi^+$.
We focus on the lepton sector for which the relevant part of the Lagrangian involving the Higgs doublets and $N_R$ is
\eq{
\label{lag:N}
-\lag\supset 
\bar{\ell}_i(Y^{(1)}_{ij}\Phi_1+Y^{(2)}_{ij}\Phi_2)e_{j R}
+\bar{N}_{R}(\la{1}_{i}\tPhi_1^\dag +\bar{N}_{R}\la{2}_{i}\tPhi_2^\dag) \ell_i
+\ums{2}\bar{N}_{R}M_RN_{R}^c+h.c.,
}
where $\ell_i$, $i=1,2,3$ are the lepton doublets. 

\begin{figure}[h]
\includegraphics[scale=0.07]{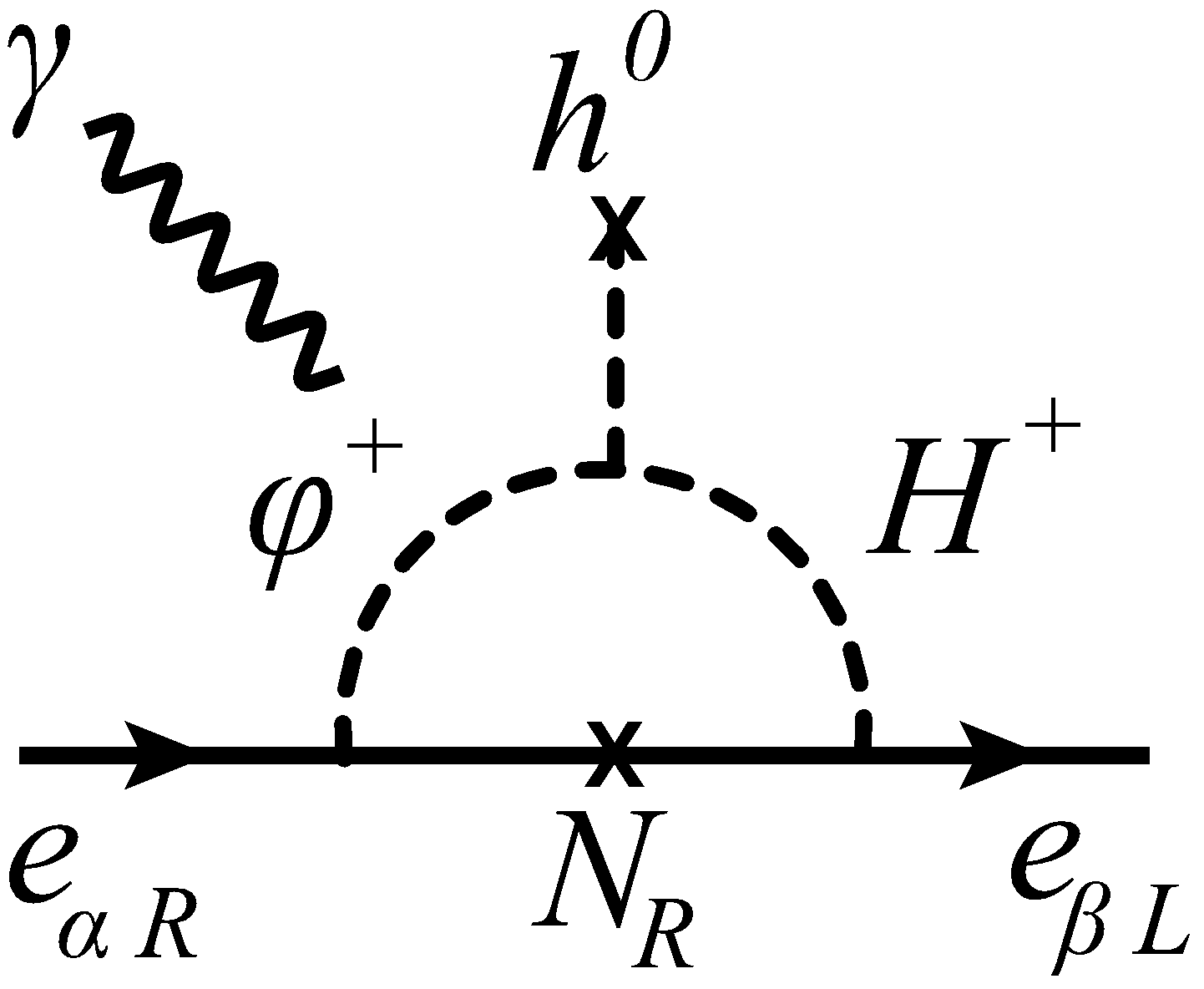}
\caption{\label{fig:g-2:NRphi}%
Chirally enhanced contribution to muon $g-2$ involving RHN $N_R$ and charged scalars.
}
\end{figure}

We seek scenarios where $a_\mu$ is explained with the participation of $N_R$ at one-loop at the 
same time that this heavy neutrino participates in the generation 
of light neutrino masses.
This contrasts with our model in Ref.\,\cite{nu.g-2} where two RHNs were present.
We require a chiral enhanced contribution to $a_\mu$ connecting $\mu_L$ with 
$\mu_R$ and then two more charged scalars ---one electroweak singlet and another residing in a 
doublet--- are necessary to close the loop as shown in Fig.\,\ref{fig:g-2:NRphi}.
From the diagram it is clear that the chiral enhancement will be proportional to the RHN Majorana 
mass $M_N$ which we require to be at most at the TeV scale.

As $\Phi_1,\Phi_2$ have the same quantum numbers, 
we can rotate them to the Higgs basis\,\cite{Georgi:1978ri,Botella:1994cs,Branco:1999fs,Davidson:2005cw}
where some physical properties become more manifest:
\eq{
\label{higgs.basis}
\mathcal{H}_1\equiv c_\beta\Phi_1+s_\beta \Phi_2\,,\qquad\quad \mathcal{H}_2=-s_\beta \Phi_1+c_\beta \Phi_2\,.
}
The vev dependent angle $\beta$ is defined by $\tan\beta=\aver{\Phi_2^0}/\aver{\Phi_1^0}$.
The doublet $\cH_1$ gets all the electroweak vacuum expectation value (vev) $v=174\,\unit{GeV}$ and carries the Goldstones bosons.

In the Higgs basis, the Lagrangian \eqref{lag:N} becomes
\eqali{
\label{yukawa:E:higgs.basis}
-\lag\supset \bar{\ell}\Big[\frac{M_e}{v}\cH_1+\Gamma_e\cH_2\Big]e_R
+\bar{N}_R\big[\La{1}\tilde{\cH}_1^\dag+\La{2}\tilde{\cH}_2^\dag\big]\ell +h.c.\,,
}
where $M_e=\diag(m_e,m_\mu,m_\tau)$.
The alignment of the original Yukawas 
\eq{
Y^{(2)}=\xi_eY^{(1)}\,,
}
leads to the alignment of the Yukawas in the Higgs basis:
\eq{
\label{alignment:Gamme}
\Gamma_e=\zeta_e\frac{M_e}{v}\,,\quad
\zeta_e\equiv\frac{\xi_e-t_\beta}{1+\xi_e t_\beta}\,.
}
Perturbativity of $\Gamma_e$ requires roughly that $|\zeta_e|\lesssim 10^3$.

For the Yukawas $\La{i}$ of the righthanded neutrino $N_R$, we similarly have
\eq{
\label{aligned:Lambda}
\La{2}=\zeta_N\La{1}\,,\quad
\zeta_N\equiv\frac{\xi_N-t_\beta}{1+\xi_N t_\beta}\,.
}
We assume $\xi_N=-1/t_\beta$ to a high degree so that $\La{1}\approx 0$ but $\La{2}$ is finite:
\eq{
\label{La2<<La1}
|\La{1}|\ll |\La{2}|\,.
}
This choice will allow us to have a low scale $N_R$ contributing to the type-I seesaw.
We then assume the following two scenarios depending if the correction to $\La{1}\approx 0$ is aligned or not to $\La{2}$.
\eqali{
\label{scenarios}
\text{$\Lambda$-aligned:}&\quad 
\text{$\La{1}$ small, cf.\,\eqref{La2<<La1}, and aligned to $\La{2}$, cf.\eqref{aligned:Lambda}};
\cr
\text{$\Lambda$-non-aligned:}&\quad
\text{$\La{1}$ small, cf.\,\eqref{La2<<La1}, and non-aligned to $\La{2}$}.
}

When we add the charged singlet scalar $\varphi^+$, we additionally have the following terms
\eq{
\label{lag:varphi}
-\lag\supset f_{i}\bar{N}_{R}e^c_{i R}\varphi^-
+\muphi\cH_2^\tp\epsilon\cH_1\varphi^-
+g_{ij}\bar{\ell}^c_i\eps\ell_j\varphi^+
+h.c.
}
We will see that one neutrino mass will be generated by tree level exchange of $N_R$ while 
another one or two masses will be generated at one-loop through one or both the Zee mechanism and Grimus-Neufeld 
mechanism. For comparison, in Ref.\,\cite{nu.g-2} we have considered two RHNs and switched off the coupling $g_{ij}$ 
above so that  neutrino masses were generated through tree level exchange of $N_R$.
Here this term is switched on and we have the Zee mechanism at action.
Due its antisymmetry, the coupling $g_{\alpha\beta}$ is inherently lepton flavor violating.

Now, let us note that the term $\cH_2^\tp\epsilon\cH_1=\Phi_2^\tp\epsilon \Phi_1$ induces a mixing between the charged higgs $H^+$ in the doublet and the scalar singlet $\varphi^+$,
\eq{
\label{phi:mixing.term}
\muphi\varphi^-\cH_2^\tp\eps\cH_1\to
-\muphi v\,\varphi^-H^+\,.
}
The orthogonal direction to $H^+$ is the charged Goldstone absorbed by the $W$.
Considering arbitrary mass terms for $H^+H^-$ and $\varphi^+\varphi^-$, the mixing term \eqref{phi:mixing.term} induces the mixing
\eq{
\mtrx{H^+\cr \varphi^+}=
\mtrx{c_\gamma & -s_\gamma\cr s_\gamma & c_\gamma}\mtrx{S_1^+\cr S_2^+}\,,
}
with angle
\eq{
\label{def.gamma}
\sin2\gamma= -\frac{2\muphi v}{M^2_{S_2}-M^2_{S_1}}\,,
}
and $S_i^+$ are the charged scalars with masses $M_{S_i}$, $i=1,2$.
We have chosen $M_{S_2}>M_{S_1}$.
The dimensionful parameter $\muphi$ cannot be arbitrarily large if its contribution to the Higgs mass 
remains natural. We require $|\muphi|\lesssim 1\,\unit{TeV}$.


We can also specify the neutral fields with definite mass in the Higgs basis \eqref{higgs.basis} as
\eq{
\cH_1=\mtrx{G^+\cr v+\frac{h_1+iG^0}{\sqrt{2}}}\,,\quad
\cH_2=\mtrx{-H^+\cr \frac{h_2+iA}{\sqrt{2}} }\,,
}
where $G^+,G^0$ are the Goldstones.
Denoting by $H$ and $h$ the CP-even mass-eigenstates, we have 
\eq{ 
\label{Hh}
\begin{pmatrix} H\\ h\end{pmatrix}=\begin{pmatrix} \cbma & \,\,\, -\sbma \\
\sbma & \,\,\,\phantom{-}\cbma \end{pmatrix}\,\begin{pmatrix} h_1 \\ 
h_2
\end{pmatrix}\,,
}
where $\mhl\leq\mhh$, $\cbma\equiv\cos(\beta-\alpha)$ and $\sbma\equiv\sin(\beta-\alpha)$ where $0\leq\beta-\alpha\leq\pi$.

Finally, we can write the Lagrangians \eqref{yukawa:E:higgs.basis} and \eqref{lag:varphi} in terms of mass definite fields\footnote{For neutrinos, this is approximate.} as
\eqali{
\label{lag:mass.basis}
-\mathscr{L}
&=
\overline{\nu}_L\Gamma_e e_{R}(c_\gamma S_1^+-s_\gamma S_2^+)+\overline{e_L}\left(\frac{M_e}{v}s_{\beta-\alpha}+\Gamma_ec_{\beta-\alpha}\right)e_{R}\frac{h}{\sqrt{2}}
\cr
&
\quad+\ \overline{e_L}\left(\frac{M_e}{v}c_{\beta-\alpha}-\Gamma_es_{\beta-\alpha}\right)e_{R}\frac{H}{\sqrt{2}}
+\frac{i}{\sqrt{2}}\overline{e}_L\Gamma_e e_{R}A
\cr
&\quad
+\ \bar{N}_R\La{2}_ke_{kL}(c_\gamma S_1^+-s_\gamma S_2^+)
\cr
&
\quad+\ \bar{N}_R\left(\Lambda^{(1)}s_{\beta-\alpha}+{\Lambda}^{(2)}c_{\beta-\alpha}\right)_k\nu_{kL}\frac{h}{\sqrt{2}}
+\bar{N}_R\left(\Lambda^{(1)}c_{\beta-\alpha}-\Lambda^{(2)}s_{\beta-\alpha}\right)_k\nu_{kL}\frac{H}{\sqrt{2}}
\cr
&
\quad+\ \frac{i}{\sqrt{2}}\bar{N}_R\Lambda^{(2)}_k\nu_{kL} A
+f_{k}^*\overline{N^c_{R}}e_{k R}(s_\gamma S_1^+ +c_\gamma S_2^+)
\cr
&
\quad+\ 2g_{kj}\overline{\nu^c_{kL}}e_{jL}(s_\gamma S_1^+ +c_\gamma S_2^+)+h.c.
}
Here, we are writing the generic interactions without assuming the Yukawa alignment condition \eqref{alignment:Gamme}.
We employ matrix notation in flavor space except for the couplings $\La{1}_k,\La{2}_k,f_k,g_{kj}$.

\section{Contributions to neutrino mass}
\label{sec:nu.mass}

Here we review the three mechanisms that contribute to neutrino masses in our setting: tree-level seesaw, 
the Grimus-Neufeld (GM) contribution and the Zee contribution.

\subsection{Tree-level mass}

Integrating out the only RHN $N_R$, we get the effective Weinberg operator:
\eq{
\label{weinberg.op}
\lag=\frac{1}{2}\bar{\ell}^c_{i}\Gamma^\tp_{i}(M_R^{-1})\Gamma_{j}
\ell_j+h.c.\,,
}
depending, in the Higgs basis, on the two Higgs doublets through
\eq{
\label{Gamma.weinberg}
\Gamma_{j}=\La{1}_{j}\tilde{\cH}_1^\dag+\La{2}_{j}\tilde{\cH}_2^\dag
\,.
}
With the presence of only one RHN, only one combination of neutrinos in the direction 
\eq{
\label{mD}
m_D=v\Lambda^{(1)}
\,,
}
will receive a mass.
The resulting rank one neutrino mass matrix at tree-level is given by
\eq{
\label{Mnu:tree}
M_\nu=-\frac{m_D^\tp m_D}{M_R}\,,
}
while the heavy neutrino remains with mass $M_N\approx M_R$.
To avoid the minus sign, we define a unit row vector $iu_0^\dag$ in the direction of $\Lambda^{(1)}$ so that 
\eq{
\label{mD:y}
m_D=iyv u_0^\dag\,,
}
where $y=|\Lambda^{(1)}|>0$.
Note that the neutrino combination that receive mass is $u_0^\dag \nu_L$, where $\nu_L$ are the neutrino components residing in the lepton doublets $\ell_\alpha$.

We seek $N_R$ at most at TeV scale that may contribute significantly to $(g-2)_\mu$.
So the necessary suppression for the Yukawa $\La{1}$, cf.\,\eqref{La2<<La1}, is analogous to the usual seesaw mechanism:
\eq{\label{eq:y_seesaw}
y\sim \frac{\sqrt{m_\nu M_R}}{v}\sim 1.8\times 10^{-6}\left(\frac{M_R}{1\,\unit{TeV}}\right)^{1/2}\,,
}
where we have taken $m_\nu\sim 0.1\,\unit{eV}$.

\subsection{Grimus-Neufeld contribution}
\label{sec:grimus.neufeld}

\begin{figure}[h]
\includegraphics[scale=1]{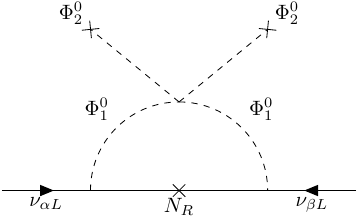}
\caption{\label{diag:GN}%
One-loop Grimus-Neufeld contribution to neutrino mass generation. 
}
\end{figure}

In the presence of two Higgs doublets and one $N_R$, the exchange of $N_R$ and neutral higgses at one-loop 
in Fig.\,\ref{diag:GN} leads to the radiative contribution of Grimus-Neufeld (GN)\,\cite{grimus.neufeld}:
\eqali{
\label{Mnu:GN}
\delta M_\nu =
-\frac{M_N}{32\pi^2}\Big\{
f_B(x_A)\Lambda_A^\tp \Lambda_A-f_B(x_H)\Lambda_H^\tp \Lambda_H-f_B(x_h)\Lambda_h^\tp \Lambda_h
\Big\}\,,
}
where $x_\phi=M^2_N/m^2_\phi$ for $\phi=A^0,H^0,h^0$, with loop function
\eq{
f_B(x)=\frac{\log x}{x-1}.
}
This function is just the Passarino-Veltman function $B_0(0,m_1,m_2)=-f_B(m_1^2/m_2^2)$.
The combination of the one-loop GN contribution \eqref{Mnu:GN} with the tree-level seesaw mass \eqref{Mnu:tree} leads to the Grimus-Neufeld \emph{mechanism}, which we distinguish from just the GN contribution.
The Yukawa couplings participating in \eqref{Mnu:GN} are combinations of the Yukawa couplings in the Higgs-flavor basis \eqref{yukawa:E:higgs.basis}:
\eqali{
\Lambda_A&=\La{2}\,,\cr
\Lambda_H&=\La{1}c_{\beta-\alpha}-\La{2}s_{\beta-\alpha}\,,\cr
\Lambda_h&=\La{1}s_{\beta-\alpha}+\La{2}c_{\beta-\alpha}\,.
}
Hence, it is clear that in the $\Lambda$-aligned case in \eqref{scenarios}, the one-loop contribution \eqref{Mnu:GN} is aligned to the tree-level contribution \eqref{Mnu:tree} and only one massive state is generated.

Higgs measurements\,\cite{ATLAS:2023qur} require that $c_{\beta-\alpha}\lesssim 0.15$ so that we are close to the alignment limit where $c_{\beta-\alpha}=0$ and $s_{\beta-\alpha}=1$.
So we assume the Higgs alignment limit for which the GN contribution \eqref{Mnu:GN} becomes
\eq{
\label{Mnu:GN:alim}
\delta M_\nu = \LGN\Lambda^{(2)\tp}\Lambda^{(2)}\,,
}
where we have defined the GN scale
\eq{
\label{GN.scale}
\LGN\equiv -\frac{M_N}{32\pi^2}[f_B(x_A)-f_B(x_H)]\,.
}
We have neglected the higgs contribution which is aligned to the tree level contribution\footnote{%
There is also the contribution from $Z$ exchange which is also aligned\,\cite{dudenas}.}
and suppressed by the small $\La{1}$ in our case.

As we seek a large contribution to $(g-2)_\mu$ from $N_R$ exchange at one-loop, we require that the Yukawa coupling $\La{2}$ be of order one and that $M_N$ be at most at TeV scale.
Consequently, the scale $\LGN$ needs to account for the smallness of neutrino masses.
As the loop factor is insufficient, we further need quasi-degenerate $A^0$ and $H^0$.
In this case, the scale \eqref{GN.scale} becomes
\eq{
\LGN\approx \frac{1}{16\pi^2}\bar{x}^{3/2}f_B'(\bar{x})(m_A-m_H)\,,
}
where $\bar{M}=(m_A+m_H)/2$ and $\bar{x}=M_R^2/\bar{M}^2$.
The function $x^{3/2}(-f_B'(x))$ is a positive and slowly varying function of $x$ with a peak $f(x)\approx 0.57$ for  $x\approx 4.48$.
The rate of decrease is slow as $f(x)\approx 0.1$ for $x\approx 0.011$ or $x\approx 5000$.
As the peak is tilted for $x>1$, it is preferable to have small $\bar{x}$ to have a small $|\LGN|$.
For $\bar{x}\sim 10^{-2}$, 
to have $\LGN\sim 0.1\,\unit{eV}$, we will need a mass difference of order $|m_A-m_H|\sim \unit{keV}$.
For generic values of parameters, this mass difference, which is much smaller than the electroweak scale, is unnaturally small compared to possible radiative corrections.
Within the 2HDM, it is possible to enforce approximate symmetries, such as in the Maximally-Symmetric-2HDM (MS-2HDM)\,\cite{dev.pilaftsis}, which protects the quasi-degeneracy. In special, the MS-2HDM has the Higgs alignment protected by symmetry as well.
Here we assume that the quasi-degeneracy of $A^0$ and $H^0$ can be protected by some approximate symmetry in our enlarged setting and explore the possibility of solving the $(g-2)_\mu$ anomaly with part of the physics that generate light neutrino masses.

\subsection{Zee mechanism}

\begin{figure}[h]
\includegraphics[scale=1.]{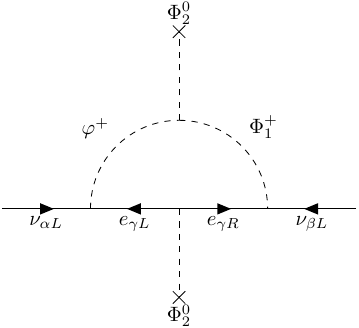}
\caption{\label{diag:zee}%
Zee mechanism of one-loop neutrino mass generation. 
}
\end{figure}

Our extension of the SM with the addition of one more Higgs doublet and one singly charged scalar naturally allows 
neutrino mass generation through the Zee mechanism\,\cite{Zee:1980ai,zee.model:2,zee.model:3} coming from the exchange 
of charged leptons and charged scalars at one-loop; see Fig.\,\ref{diag:zee}.
This contribution is given by\,\cite{herrero}
\eq{
\label{Mnu:Zee}
\delta M_\nu = \Azee g M_e\Gamma_e^\dag+(~)^\tp 
\,,
}
where $M_e=\diag(m_e,m_\mu,m_\tau)$, $(~)^\tp$ denotes the transpose of the previous term and 
\eq{
\label{def:AZee}
\Azee\equiv \frac{s_{2\gamma}}{16\pi^2 t_\beta}\log\frac{M_{S_2}^2}{M_{S_1}^2}\,.
}
In the loop function, we have neglected corrections in $(\log x)/x$ where $x=M^2_{S_i}/m_\alpha^2$.
The mixing angle $\gamma$ is given in Eq.\,\eqref{def.gamma} and $M_{S_{1,2}}$ are the masses of the charged scalars.
The Yukawa coupling $\Gamma_e$ is given in \eqref{yukawa:E:higgs.basis} while the 
matrix coupling $g$ is given in \eqref{lag:varphi}.
Due to the antisymmetry of $g$, it is automatic that $u_{\rm Zee}^\tp\delta M_\nu u_{\rm Zee}=0$ 
where $u_\text{Zee}=(g_{\mu\tau},-g_{e\tau},g_{e\mu})$\,\cite{herrero:singly.charged}.

When we consider the Yukawa alignment \eqref{alignment:Gamme}, the Zee mechanism to neutrino mass \eqref{Mnu:Zee} becomes
\eq{
\label{Mnu:Zee:yuk.a}
\delta M_\nu = \frac{\Azee\zeta_e}{v}(g M^2_e-M_e^2g)
\,.
}
This form with zeros in the diagonal is proportional to the Zee-Wolfenstein version\,\cite{Zee:1980ai,wolfenstein} and 
is 
known\,\cite{zee-wolfenstein:out} to be incompatible with actual neutrino data.
So \eqref{Mnu:Zee:yuk.a} cannot be the dominant contribution to neutrino mass generation.
In Sec.\,\ref{sec:combining} we will combine the Zee mechanism with the tree-level seesaw and the GN contribution to generate realistic neutrino masses.

Barring accidental cancellations, to obtain $\delta M_\nu\sim 0.1\,\unit{eV}$, we will need the $g_{\alpha\beta}$ components to be of order
\eq{
\label{AZee:typical}
\Azee\zeta_e g_{\alpha\beta}\sim 0.1\,\unit{eV}\times \frac{v}{m_\tau^2}\sim 5\times 10^{-9}\,.
}
It is clear that the typical values for the couplings $\zeta_e g_{\alpha\beta}$ should be at most $10^{-4}$.
For example, $A=6.3\times 10^{-6}$ for $t_\beta=100, s_{2\gamma}=0.1$ and neglecting the logarithm.

\section{Combining different mechanisms}
\label{sec:combining}

Given that the tree-level seesaw mechanism in \eqref{Mnu:tree} is only capable of generating one light neutrino mass, we need to explore the combination with one or both the GN contribution or Zee contribution.
It should be reminded that the GN mechanism in its general form\footnote{%
With one RHN and two Higgs doublets.}
---combining the tree-level seesaw and the one-loop GN contribution---   is capable of generating a sufficiently general neutrino mass matrix giving mass to two neutrino states.
But here we search for models where $N_R$ is capable of generating an one-loop contribution sufficient to explain the 
$(g-2)_\mu$ deviation from the SM, cf.\,\eqref{eq:amuBSM}, and yet avoiding the possible CLFV constraints.
Another point worth emphasizing is that when the two Yukawas $\La{1}$ and $\La{2}$ are aligned, the one-loop contribution \eqref{Mnu:GN} becomes aligned with the tree-level contribution \eqref{Mnu:tree} and then the GN mechanism is not sufficient for neutrino mass.

\subsection{Seesaw and Grimus-Neufeld}

Let us first consider the case of suppressed couplings for $g$ so that the contribution from the Zee mechanism is subdominant and we can consider the tree level \eqref{Mnu:tree} and GN \eqref{Mnu:GN} contributions to neutrino mass. 
This possibility is usually referred to as the GN mechanism (or model) and is only possible in the $\Lambda$-non-aligned scenario \eqref{scenarios}.

Considering the rank one Dirac mass matrix in \eqref{mD:y}, the neutrino mass matrix 
at tree level is nonzero only in the space $\nu_{0}=u_0^\dag\nu_L$ and $N_R^c$.
For NO (IO) $\nu_0=\nu_3$ ($\nu_0=\nu_2$).
Instead of the seesaw formula \eqref{Mnu:tree}, here we consider the exact diagonalization before including
the one-loop contribution.
In this space, the $2\times 2$ neutrino mass matrix is
\eq{
\label{M.eff:tree}
\mtrx{0 & iyv\cr iyv & M_R}\,,
}
with masses
\eqali{
m_0&=\sqrt{\ums{4}M_R^2+(yv)^2}-\ums{2}M_R\approx \frac{(yv)^2}{M_R}\,,
\cr
m_4&=\sqrt{\ums{4}M_R^2+(yv)^2}+\ums{2}M_R\approx M_R\,,
}
where the last approximations are valid for $M_R\gg |m_D|$.
The inverse relations are $yv=|m_D|=\sqrt{m_0m_4}$ and $M_R=m_4-m_0$.
The exact diagonalization matrix for the seeasaw is\,\cite{stockinger.22}
\eq{
\label{ss.matrix}
\mtrx{c_{ss} & -is_{ss}\cr -is_{ss} & c_{ss}}\,,
\quad
\text{with ~~}
c_{ss}=\sqrt{\frac{m_4}{m_4+m_0}}\,,~~
s_{ss}=\sqrt{\frac{m_0}{m_4+m_0}}\,.
}

With the one-loop GN contribution \eqref{Mnu:GN:alim}, another combination gets mass.
Let us call the additional direction $u_\perp$ and define it as a unit vector in the space ${\la{1}}^\dag,{\la{2}}^\dag$ orthogonal to $u_0$.
The coupling $\Lambda^{(2)}$ then has only components in these two directions: 
\eq{
\label{Lambda2:dd'}
\Lambda^{(2)}=\Lambda^{(2)}_\perp u_\perp^\dag +\Lambda^{(2)}_0u_0^\dag
= \pm de^{+i\alpha/2} u_\perp^\dag+ d'e^{-i\alpha/2}u_0^\dag
\,.
}
To obtain the last expression, we choose the global phase of $u_\perp$ appropriately so that $d,d'$ are real positive.
The possible sign is due to $\sign(\LGN)=\pm 1$ defined in \eqref{GN.scale}.
Our definition for the parameters $d,d'$ differ from \cite{grimus.neufeld} and from \cite{stockinger.22,dudenas} as well.
We denote the neutrino field in the direction of $u_\perp$ as $\nu_{\perp}=u_\perp^\dag \nu_L$.
Then the GN mechanism will induce a nonzero mass contribution to the neutrinos $(\nu_{\perp},\nu_{0})$.

Taking the tree-level contribution \eqref{M.eff:tree} and neglecting the small angle seesaw rotation \eqref{ss.matrix}, we obtain the mass matrix in the space $(\nu_\perp,\nu_0,N_R^c)$:
\eq{
\left(
\begin{array}{c|c}
\Sigma & 0\cr 
\hline
0 & m_4
\end{array}
\right)\,,
}
with the effective $2\times 2$ matrix given by
\eq{
\label{Sigma:GNM}
\Sigma \approx \mtrx{0&0\cr 0&m_0}
	+ \mtrx{\LGN d^2e^{i\alpha} & |\LGN dd'|\cr |\LGN dd'| & \LGN d'^2e^{-i\alpha}}\,.
}
The second contribution is due to GN; cf.\,\eqref{Mnu:GN:alim}.
The parametrization in \eqref{Lambda2:dd'} was motivated to obtain a real off-diagonal entry in $\Sigma$ so that the analytic diagonalization in appendix \ref{ap:2x2} of a $2\times 2$ Majorana matrix could be used directly.
Appendix \ref{ap:masses} shows how to use the masses as input parameters.
The diagonalization yields
\eq{
\label{Sigma:diag}
\text{(NO)}\quad
U_\Sigma^\tp\Sigma U_\Sigma=\diag(\hat{m}_2,\hat{m}_3)\,,
\quad
\text{(IO)}\quad
U_\Sigma^\tp\Sigma U_\Sigma=\diag(\hat{m}_1,\hat{m}_2)\,,
}
where the hatted masses are one-loop corrected (pole) masses.
Note that $\Sigma$ only depends on four parameters ($m_0,\sqrt{\LGN} d,\sqrt{\LGN} d',\alpha$) and it is not a general $2\times 2$ symmetric matrix. 
In appendix \ref{ap:GMN} we show how to use the two masses in \eqref{Sigma:diag}, $y$ and $c=|\LGN dd'|$ as input parameters.

Including now the fourth neutrino $\nu_\star$ orthogonal to $(\nu_\perp,\nu_0,N_R^c)$ which remains massless at one-loop, we relate the $3\times 3$ mixing matrix of the light neutrinos to the PMNS matrix $V$:
\eq{
\label{V:U0}
\text{(NO)}\quad
V=U_0
\left(
\begin{array}{c|c}
1 & 0\cr 
\hline
0 & U_\Sigma
\end{array}
\right)\,,
\quad
\text{(IO)}\quad
V=U_0
\left(
\begin{array}{c|c}
U_\Sigma & 0 \cr
\hline
0 & 1
\end{array}
\right)\,,
}
where
\eq{
\label{U0}
\text{(NO)}\quad
U_0=\Big(\star\Big| u_\perp\Big| u_0\Big)\,,
\quad
\text{(IO)}\quad
U_0=\Big(u_\perp\Big| u_0\Big| \star\Big)\,,
}
is the matrix of basis change to $(\nu_{\star},\nu_\perp,\nu_0)$ for NO or analogously for IO.
Since this matrix is a generic unitary matrix, we can invert \eqref{V:U0} and obtain
\eq{
\label{U0:inversion}
\text{(NO)}\quad
U_0=V\left(
\begin{array}{c|c}
1 & 0\cr 
\hline
0 & U_\Sigma^\dag
\end{array}
\right)\,,
\quad
\text{(IO)}\quad
U_0=V\left(
\begin{array}{c|c}
U_\Sigma^\dag & 0 \cr
\hline
0 & 1
\end{array}
\right)\,.
}
The $2\times 2$ matrix $U_\Sigma$ is not completely free and its parameters are restricted to give the correct masses in \eqref{Sigma:diag}.
For one massless neutrino, there is only one physical Majorana phase and the PMNS matrix can be written as
\eq{
\text{(NO)}\quad
V=V_0\mtrx{1&&\cr &1&\cr &&e^{i\alpha'}}\,,
\quad
\text{(IO)}\quad
V=V_0\mtrx{1&&\cr &e^{i\alpha'}&\cr &&1}\,,
}
where $V_0$ is in the standard parametrization and $\alpha'$ is the physical Majorana phase.

The relation \eqref{U0:inversion} implies that one of the columns of $U_0$ coincides with the respective column of the 
PMNS matrix and, from unitarity of $U_0$, the components of the columns containing $u_0$ and $u_\perp$ are also 
constrained as
\eqali{
\label{u0+uperp}
\text{(NO)}&\quad
|u_{\perp \alpha}|^2+|u_{0 \alpha}|^2=1-|V_{\alpha 1}|^2\approx (0.3,0.9,0.8)\,,
\cr
\text{(IO)}&\quad
|u_{\perp \alpha}|^2+|u_{0 \alpha}|^2=1-|V_{\alpha 3}|^2\approx (0.98,0.6,0.5)\,.
}

We can now analyze $\La{2}_e$ and $\La{2}_\tau$ in \eqref{Lambda2:dd'} which, if unsupressed, will contribute unacceptably to CLFV processes such as $\mu\to e\gamma$ and $\tau\to \mu\gamma$, respectively.
On the other hand, we need a large $\La{2}_\mu$ to contribute to $a_\mu$.
To have an order one $\La{2}_\mu$, it is clear that we cannot suppress both $d,d'$.
Due to \eqref{u0+uperp}, the $e$- and $\tau$-components of $u_0$ and $u_\perp$ cannot be simultaneously suppressed as well.
Hence a cancellation between the two terms in \eqref{Lambda2:dd'} is necessary for $\alpha=e,\tau$.
The cancellation for $\La{2}_e,\La{2}_\tau$ requires
\eq{
\label{uperp=ru0}
u_{\perp e}=ru_{0e}\,,\quad
u_{\perp \tau}=ru_{0\tau}\,
\,,
}
with the common factor
\eq{
r\equiv -\sign(\Lambda)\frac{d'e^{+i\alpha}}{d}=-\frac{\Sigma_{12}}{\Sigma_{11}^*}
\,,
}
where in the second equality we used \eqref{Sigma:GNM}.
We will show that the requirement \eqref{uperp=ru0} is incompatible with \eqref{u0+uperp} and the orthonormality of $\{u_0,u_\perp\}$.

Let us take NO for definiteness.
Imposing \eqref{uperp=ru0} to \eqref{u0+uperp} implies on the one hand that
\eq{
\label{u0+uperp:2}
|u_{0e}|^2+|u_{0\tau}|^2=\frac{1+|V_{\mu 1}|^2}{1+|r|^2}\,.
}
On the other hand, $|u_0|^2=1$, $|u_\perp|^2=1$ and $u_0^\dag u_\perp=0$ lead respectively to
\eqali{
|u_{0\mu}|^2&=1-(|u_{0e}|^2+|u_{0\tau}|^2)\,,
\cr
|u_{\perp\mu}|^2&=1-|r|^2(|u_{0e}|^2+|u_{0\tau}|^2)\,,
\cr
u_{0\mu}^*u_{\perp\mu}&=-r(|u_{0e}|^2+|u_{0\tau}|^2)\,,
}
after using \eqref{uperp=ru0}.
Combination of these relations leads to
\eq{
|u_{0e}|^2+|u_{0\tau}|^2=\frac{1}{1+|r|^2}\,,
}
which is clearly incompatible with \eqref{u0+uperp:2} with $|V_{\mu 1}|^2\approx 0.1$.
The conclusion is the same for IO.

We conclude that in the GN mechanism with the known PMNS mixing, it is not possible to have simultaneously $\La{2}_e=\La{2}_\tau=0$ and $\La{2}_\mu=O(1)$.
Given the incompatibility above, even suppressing $\La{2}_e,\La{2}_\tau$ sufficiently to evade CLFV constraints is not 
possible.
The required small values for $\La{2}_e,\La{2}_\tau$ are discussed in Sec.\,\ref{sec:results:non}.

\subsection{Seesaw, Grimus-Neufeld and Zee}

As the Grimus-Neufeld mechanism is not capable of simultaneously generating the light neutrino masses,
evading CLFV and explaining $(g-2)_\mu$, we switch on the coupling $g\ell\ell\varphi^+$ and consider the one-loop 
Zee contribution for neutrino mass.

\subsubsection{Aligned case}
\label{sec:aligned}

Here we consider the $\Lambda$-aligned case in \eqref{scenarios}, where the neutrino Yukawas $\La{2},\La{1}$ are proportional.
In this case, the seesaw and GN contribution are aligned and only one neutrino gets a mass at one-loop.
Then the addition of the Zee mechanism is mandatory for neutrino mass.
The contribution to light neutrino masses of the tree-level seesaw \eqref{Mnu:tree} and the GN contribution \eqref{Mnu:GN:alim} is rank one and proportional to ${\La{1}}^\tp\La{1}\propto {\La{2}}^\tp\La{2}$.
As we will see in Sec.\,\ref{sec:results}, $\La{2}_e$ will lead to a chiral enhanced contribution to $\mu\to e\gamma$ and we need to greatly suppress this coupling.
For simplicity, we take $\La{2}_e=0$.
Then the tree-level seesaw and GN contribution can be written as
\eq{
\label{L-aligned:tree+GNM}
M_\nu\Big|_{\rm tree+GN}=\mtrx{0 & 0 & 0\cr &a^2&ab \cr &&b^2}\,,
}
where $a,b$ can be taken real by rephasing fields.
This block diagonal structure obviously cannot generate the correct PMNS mixing.

To be able to obtain a realistic neutrino mass matrix, we add to \eqref{L-aligned:tree+GNM} the contribution from the Zee mechanism in \eqref{Mnu:Zee:yuk.a} which leads to the structure
\eq{
\label{Mnu:ss.zee:yuk.a}
M_\nu=\mtrx{0 & 0 & 0\cr &a^2&ab \cr &&b^2}+\mtrx{&h&0\cr \star&&f\cr \star&\star&}
=\mtrx{0&C&0\cr C&A&D\cr 0&D&B}
\,.
}
We adopt vanishing $(\mu\tau)$ entry in the second matrix in the middle equation (Zee) by taking $g_{e\tau}=0$ to solve the Cabibbo angle anomaly in Sec.\,\ref{sec:cabibbo}.
Then we can take real $h$ while $f$ is complex. 
We have a total of 5 parameters to describe the neutrino mass matrix, which is the same number of observables available: three angles $\theta_{ij}$ and two mass differences $\Delta m^2_{ij}$.

The structure \eqref{Mnu:ss.zee:yuk.a} is a two-zero texture extensively studied in the literature since
Ref.\,\cite{2-texture}.
This is, e.g., texture A2 in Ref.\,\cite{2-texture}, for which only NO is feasible\,\cite{kumar.texture}.
For the current best-fit values\,\cite{nufit.24} of $\theta_{ij}$ and $\Delta m^2_{ij}$, 
solutions can be easily found, yielding
\eq{
\label{aligned:A.B}
A=28.02\,\unit{meV}\,,\quad
B=23.33\,\unit{meV}\,,\quad
C=9.99\,\unit{meV}\,,\quad
D=23.63\,\unit{meV}\,,
}
for the lightest mass $m_1=4.72\,\unit{meV}$
and CP phases $\delta=1.413, \alpha=-1.394, \beta=-0.745$.
So we conclude that $a\sim b$ are of the same order.
The same is true for $\La{2}_\mu\sim \La{2}_\tau$.
While $\La{2}_\mu= O(1)$ allows us to solve $a_\mu$, order one $\La{2}_\tau$ leads to 
too large contributions to $\tau\to \mu\gamma$; see Sec.\,\ref{sec:results:aligned}.

\subsubsection{Non-aligned case}
\label{sec:non-aligned}

In case $\La{1}$ and $\La{2}$ are non-aligned, we can choose $\La{2}=(0,\La{2}_\mu,0)$ to turn-off the couplings to the $e$ and $\tau$ flavors.
Then, we have the structure
\eq{
\label{Mnu:ss+gn+zee}
M_\nu=\mtrx{a^2 &ab&ac \cr &b^2+d^2&bc \cr &&c^2}+\mtrx{&h&e\cr h&&f\cr e&f&}
=
\mtrx{A & D& E \cr & B & F \cr &&C}
\,,
}
where $(a,b,c)=(-i)M_R^{-1/2}v(\La{1}_e,\La{1}_\mu,\La{1}_\tau)$, $d^2=\LGN(\La{2}_\mu)^2$ and
\eqali{
\label{efh}
e&=\frac{\Azee\zeta_e}{v}g_{e\tau}m_\tau^2=0.1\,\unit{eV}(1300\times\zeta_eg_{e\tau})\frac{\Azee}{10^{-5}}\,,
\cr
f&=\frac{\Azee\zeta_e}{v}g_{\mu\tau}m_\tau^2=0.1\,\unit{eV}(1300\times\zeta_eg_{\mu\tau})\frac{\Azee}{10^{-5}}\,,
\cr
h&=\frac{\Azee\zeta_e}{v}g_{e\mu}m_\mu^2= 
0.1\,\unit{eV}(1300\times\zeta_eg_{e\mu})\frac{\Azee}{10^{-5}}\,.
}

\begin{figure}[h]
\includegraphics[scale=.37]{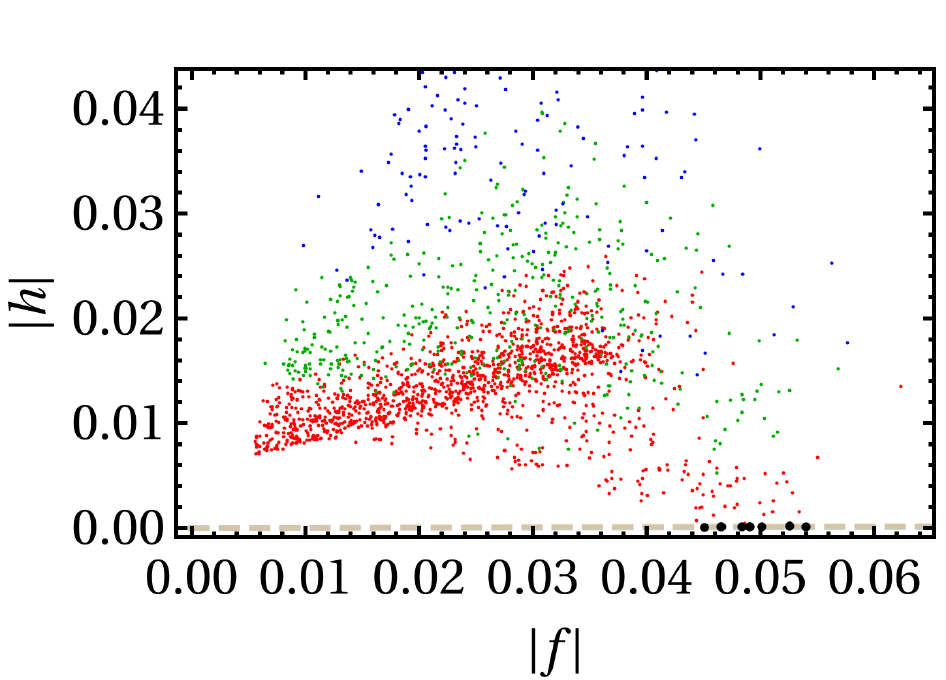}
\includegraphics[scale=.6]{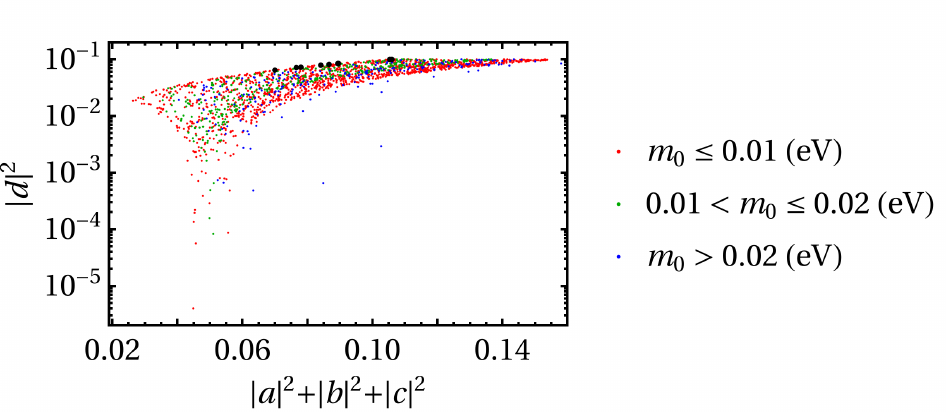}
\caption{\label{fig:hxf}
Left: Scatter plot of $|h|$ (eV) against $|f|$ (eV) in \eqref{Mnu:ss+gn+zee} for different values of the minimal mass $m_0=m_1$ (NO) when $e=0$ in \eqref{Mnu:ss+gn+zee}.
Right: 
Scatter plot of $|d^2|$ (eV) against $|a|^2+|b|^2+|c|^2$  (eV) in \eqref{Mnu:ss+gn+zee} for different values of the minimal mass $m_0=m_1$ (NO) when $e=0$ in \eqref{Mnu:ss+gn+zee}. 
}
\end{figure}

Due to Cabibbo angle anomaly and lepton universality constraints to be considered in 
Sec.\,\ref{sec:contraints},
we will consider the case where $e=0$ ($g_{e\tau}=0$).
Hence the following vanishing minor condition holds:
\eq{
\label{minor}
(M_\nu)^{-1}_{22}\sim AC-E^2=0\,.
}
In this case only NO is possible.
We show in Fig.\,\ref{fig:hxf} the scatter plot for $(|f|,|h|)$ (left panel) when we fix the neutrino mass differences
$\Delta m^2_{12},\Delta m^2_{23}$ and the mixing angles $\theta_{12},\theta_{23},\theta_{13}$ to the
experimental values within $3\sigma$\,\cite{nufit.24}.
We take $m_1,|d|^2\in [0,0.1\,\unit{eV}]$.
We see that $|f|,|h|$ fall in the range of the tens of the meV, compatible with the overall neutrino mass scale.
In the right panel we also show the scatter plot for $|d|^2$ against $|a|^2+|b|^2+|c|^2$ which also are distributed
around the typical values of neutrino masses.

We can still analyze some generic consequences.
From \eqref{Mnu:ss+gn+zee}, it is clear that
\eq{
\label{h/f:g}
\frac{|h|}{|f|}=\frac{m^2_\mu}{m^2_\tau}\left|\frac{g_{e\mu}}{g_{\mu\tau}}\right|
=3.5\times 10^{-3}\left|\frac{g_{e\mu}}{g_{\mu\tau}}\right|
\,.
}
Therefore, if $|g_{e\mu}|\sim |g_{\mu\tau}|$ as required to solve the Cabibbo anomaly, we must have 
$|h|\ll |f|$.
By requiring
\eq{
\label{h/f:range}
1.22\times 10^{-3}<\frac{|h|}{|f|}<2.89\times 10^{-3}\,,
}
we have found 9 points which are depicted as black points in Fig.\,\ref{fig:hxf}.
For these black points, 
\eq{
\label{range:m1.f}
m_1\in [5.2,6.6]\,\unit{meV},\quad
|f|\in [44,54]\,\unit{meV}\,.
}
We expect that the inclusion of more points will only lead to a mild widening of these ranges.
The specific limits in \eqref{h/f:range} will be discussed in Sec.\,\ref{sec:lepton.univ}.

\section{Flavor and electroweak constraints}
\label{sec:contraints}

Here we discuss the various contributions to CLFV processes, lepton universality violation (LUV) 
and electroweak constraints.

\subsection{Dipole moments and $\ell\to \ell'\gamma$}
\label{sec:dipole}

Using an effective theory approach, the operator relevant to lepton dipole moments 
and charged lepton flavor violation (CLFV) is the dipole operator
\eqali{
\label{L.eff:photon}
\lag_{\text{dipole}}&=
-\left(C^{\sigma R}_{\alpha\beta}\bar{e}_{\alpha L}\sigma_{\mu\nu}e_{\beta R} F^{\mu\nu}
+h.c.\right)
\,.
}
It is well known that the dipole operator is chirality flipping and chiral enhancement is possible within our model.

The Wilson coefficient of the dipole operator
contributes to the dipole moments and $\ell_\alpha\to \ell_\beta\gamma$ as
\eqali{
\label{a.mu:C}
a_{\alpha}&=-\frac{4 m_\alpha}{e}\re(C^{\sigma R}_{\alpha\alpha})\,,
\cr
d_{\alpha}&=-2\im(C^{\sigma R}_{\alpha\alpha})\,,
\cr
\Br[\ell_\alpha\to \ell_\beta\gamma] &=\frac{m^3_\alpha}{4\pi \Gamma_\alpha}(|C^{\sigma 
R}_{\alpha\beta}|^2+|C^{\sigma R}_{\beta\alpha}|^2)\,,
}
where $a_\alpha,d_\alpha$ are the contributions to the magnetic and electric dipole moments, 
respectively.
These formulas assume that our covariant derivative in QED is $D_\mu=\partial_\mu+ieQA_\mu$ and 
similarly for the SM and beyond.

In the Higgs basis, the complete contribution from the diagram in Fig.\,\ref{fig:g-2:NRphi} with 
exchange of charged scalars is
\subeqali[C.dipole]{
\label{cR:lambda.f}
\frac{16\pi^2}{e}C^{\sigma R}_{\beta\alpha}
&=
c_\gamma s_\gamma \frac{{\Lambda^{(2)}_{\beta }}^\dag f^*_{\alpha}}{M_{N}}\left[x_{2}f_S (x_{2})-x_{1}f_S (x_{1})\right]\,,
\\
\label{cR:lambda.lambda}
&\quad +\ 
m_\alpha \Lambda^{(2)\dag}_{\beta }
    \left[
    \frac{c_\gamma^2}{M^2_{S_1}}\tf_S(x_{1})
    +\frac{s_\gamma^2}{M^2_{S_2}}\tf_S(x_{2})
    \right]
\Lambda^{(2)}_{\alpha}
\\
\label{cR:f.f}
&~~\ 
+\
m_\beta  f_{\beta }
    \left[
    \frac{s_\gamma^2}{M^2_{S_1}}\tf_S(x_{1})
    +\frac{c_\gamma^2}{M^2_{S_2}}\tf_S(x_{2})
    \right]
f_{\alpha}^{*}
\,,
\\
\label{cR:g.g}
&~~\ 
+\
\frac{m_\alpha}{6} \sum_i g_{i\beta}^* g_{i\alpha}
    \left[
    \frac{s_\gamma^2}{M^2_{S_1}}
    +\frac{c_\gamma^2}{M^2_{S_2}}
    \right]
\,,
\\
\label{cR:Gamma.Gamma}
&~~\ 
+\
\frac{m_\beta}{24} \sum_i (\Gamma_e)_{i\beta}^* (\Gamma_e)_{i\alpha}
    \left[
    \frac{c_\gamma^2}{M^2_{S_1}}
    +\frac{s_\gamma^2}{M^2_{S_2}}
    \right]\,,
}
where $x_{k}\equiv M^2_{N}/M^2_{S_k}$
and the loop functions are\,\cite{crivellin:g-2,calibbi.ziegler}
\eqali{
f_S (x)&\equiv \frac{x^2-1-2x \log x}{4(x-1)^3}\,,
\cr
\tf_S(x)&\equiv \frac{2x^3+3x^2-6x+1-6x^2\log x}{24(x-1)^4}\,.
}
We neglect contributions suppressed by the $\nu{-}N$ mixing in the seesaw.

Let us discuss the various contributions.
The first contribution \eqref{cR:lambda.f} is the chirally enhanced (left-right) contribution 
while the rest of them are not chirally enhanced (left-left or right-right respectively) as the 
chiral flip comes from the external lines.
So for $a_\mu$ the chirally enhanced contribution is larger than the non-chirally enhanced 
contributions by a 
factor
\eq{
c_\gamma s_\gamma \frac{M_S^2}{m_\mu M_N}\sim c_\gamma s_\gamma\times 10^4\,,
}
for order one couplings, $M_N\sim 10\,\unit{GeV}$ and $M_S\sim 100\,\unit{GeV}$.
For these values, if we want the chiral enhanced contribution to dominate, we require $|s_\gamma|\gtrsim 10^{-3}$.
Given the definition of the angle $\gamma$ in \eqref{def.gamma}, the trilinear coupling $\mu_\varphi$ also 
cannot be arbitrarily small if we require the chirally enhanced contribution to dominate:
\eq{
\label{muphi>}
|\mu_\varphi|\gtrsim 0.1\,\unit{GeV}\times\frac{M_N}{10\,\unit{GeV}}
\,.
}
The dominance of the chirally enhanced contribution is important because the non-chirally enhanced 
contributions are positive definite and leads to a contribution to $a_\mu$ which is negative 
definite, contrary to the experimental observation.
They are similar to the contribution of $N_R$ exchange in the usual seesaw 
models\,\cite{freitas,coy.frigerio}.

In Table \ref{tab:LFV} we show the current and future limits for different $\ell_\alpha\to 
\ell_\beta\gamma$.
Limits for other CLFV processes are also shown.
\begin{table}[h]
\[
\begin{array}{|c|c|c|c|}
\hline  \text{Observable}    & \text{Current limit} & \text{Future limit} 
    \cr
\hline
\text{Br}(\mu\to eee)       &      <1.0 \times 10^{-12} \text{\cite{pdg}}       &  10^{-16} 
\text{\cite{mu3e}} \cr
\text{Br}(\tau\to \mu\mu\mu)       &   <2.1 \times 10^{-8} \text{\cite{HFLAV:2019otj}}    & 
3.4\times 10^{-10} \text{\cite{belle2:book}} \cr
\text{Br}(\tau\to \mu ee)       &  <8.4 \times 10^{-9} \text{\cite{HFLAV:2019otj}}  & 2.9\times 
10^{-10} \text{\cite{belle2:book}}\cr   
\text{Br}(\tau\to eee)       &  <1.4 \times 10^{-8} \text{\cite{HFLAV:2019otj}}  & 4.3\times 
10^{-10} \text{\cite{belle2:book}}\cr
\text{Br}(\tau\to e\mu\mu)       &  <1.6 \times 10^{-8} \text{\cite{HFLAV:2019otj}} & 4.3\times 
10^{-10} \text{\cite{belle2:book}}\cr
\hline
\text{Br}(\mu\to e\gamma)       &  <1.5 \times 10^{-13} \text{\cite{meg2:result}} & 6\times 
10^{-14}\text{\cite{meg2}} \cr
\text{Br}(\tau\to \mu\gamma)       &  <4.2 \times 10^{-8} \text{\cite{belle:tau.mu.gamma}} & 10^{-9} 
\text{\cite{belle2:book}} \cr
\text{Br}(\tau\to e\gamma)       &  <3.3 \times 10^{-8} \text{\cite{pdg}} & 3\times 10^{-9} 
\text{\cite{belle2:book}} \cr 
\hline
\Gamma^{\rm conv}_{\mu \rightarrow e}/\Gamma^{\rm capt}_{N} & <7.0 \times 10^{-13} 
\text{\cite{SINDRUMII:2006dvw}}^{*} & 3 \times 10^{-17} \text{\cite{mu2e,comet}}^{**}/ 
10^{-18}\text{\cite{jparc}}^{\dagger}\cr
\hline
\end{array}
\]
\caption{\label{tab:LFV}
Current and future limits for charged lepton flavor violating processes at 90\% CL.
For $\mu e$ conversion, the nucleus is Au (*) and Al (**) or Ti ($\dagger$). 
}
\end{table}

For our model in Sec.\,\ref{sec:non-aligned} with non-aligned $\La{1}$ and $\La{2}$, and with 
$g_{e\tau}=0$, only the decay $\tau\to e\gamma$ is relevant with branching ratio proportional to 
$|g_{\mu\tau}g_{\mu e}|^2$.
The limits for $\mu e$ conversion in different nuclei\,\cite{mu-e:conversion} are expected to significantly improve in 
the future but they do not apply as well.

\subsection{$\ell\to \ell'\ell''\ell'''$}

For generic 2HDMs the coupling of the neutral scalars to charged leptons may violate family lepton 
numbers which leads to flavor changing decays of charged leptons.
In our setting, cf.\,\eqref{lag:mass.basis}, this is absent because of Yukawa alignment and the coupling $\Gamma_e$ is diagonal.
Lepton flavor violating Higgs decay is not possible for the same reason.
Other flavor changing tree-level decays with exchange of $N$ or $S_i^+$ are equally not possible because 
only $\La{2}_\mu$ and $f_\mu$ are nonzero.

At one-loop, flavor changing decay of a charged lepton to three charged leptons becomes possible.
In our case, only the coupling $g_{\alpha\beta}$ is flavor changing and we will consider the case where $g_{e\tau}=0$;
see around \eqref{crivellin.fit}.
This only allows the transition $\tau\to e$ while $\mu\to e$ and $\tau\to \mu$ are not possible.
So for $\tau\to e(\ell^+\ell^-)$, $\ell=e,\mu$, there are contributions of on-shell and off-shell photonic penguin
operators similar to the contribution to $\tau\to e\gamma$.
Usually the on-shell contribution dominates over the off-shell contribution\,\cite{kuno.okada}.
Without additional contributions, the branching for $\tau\to e(\ell^+\ell^-)$ can be
directly related to the branching for $\tau\to e\gamma$ and the latter is automatically more
stringent\,\cite{herrero:singly.charged}.

For large $g_{\alpha\beta}$, box diagram contributions become relevant.
Considering $g_{e\tau}=0$, these box contributions lead to
\eqali{
\Br[\tau\to e\mu\mu]&=\frac{m_\tau^5}{6\pi^3\Gamma_\tau}
\left|\frac{g_{e\mu}^*g_{\mu\tau}(|g_{e\mu}|^2+|g_{\mu\tau}|^2)}{64\pi^2\bar{M}^2}\right|^2
\,,
\cr
\Br[\tau\to eee]&=\frac{m_\tau^5}{3\pi^3\Gamma_\tau}
\left|\frac{g_{e\mu}^*g_{\mu\tau}|g_{e\mu}|^2}{64\pi^2\bar{M}^2}\right|^2
\,,
}
where
\eq{
\frac{1}{\bar{M}^2}\equiv \frac{c^4_\gamma}{M_{S_2}^2}+\frac{2c_\gamma^2 s_{\gamma}^2}{M_{S_2}^2-M_{S_1}^2}
\log\frac{M_{S_2}^2}{M_{S_1}^2}+\frac{s^4_\gamma}{M_{S_1}^2}\,.
}
We have adapted the results of Refs.\,\cite{crivellin:singly.charged,mituda} to our case.
The photonic on and off shell contribution for one singly charged scalar can be found in 
Ref.\,\cite{crivellin:singly.charged}
and they are subdominant for large $g_{\alpha\beta}$.
For future convenience we also define the ratio
\eq{
\label{def:M'S}
M'_{S}\equiv \frac{\tM_2^2}{\bar{M}}\,.
}

Finally, the box diagram with $\Gamma_e$ (flavor conserving) and $g_{ij}$ (flavor changing) is not
possible in the same fermion line because of chirality mismatch.
We need chiral flipping which in this case is proportional to the light neutrino mass.

\subsection{Cabibbo angle anomaly}
\label{sec:cabibbo}

By tree-level exchange of the charged scalars $S_i^+$, there is an additional contribution to 
muon decay which leads to a modification to the Fermi constant\,\cite{herrero,nebot}.
The leading contribution, after Fierz reorderings, has the same contribution as $W$ exchange 
in the operator\,\cite{jenkins.manohar:match} $(\bar{\nu}_{L\mu}\gamma^\mu\nu_{Le})(\bar{e}_L\gamma_\mu\mu_L)$
modifying the Fermi constant $G_F=G_\mu$ as
\eq{
\label{GF.smeft}
\frac{4G_\mu}{\sqrt{2}}=
\frac{4G_F^{\rm SM}}{\sqrt{2}}
-C_{\underset{\mu ee\mu}{ll}}
-C_{\underset{e\mu\mu e}{ll}}
+2C^{(3)}_{\underset{\mu\mu}{Hl}}
+2C^{(3)}_{\underset{ee}{Hl}}\,.
}
This deviation is induced by the Wilson coefficients of the SMEFT operators
\eqali{
\label{smeft.op}
Q_{ll}&=(\bar{l}_p\gamma_\mu l_r)(\bar{l}_s\gamma^\mu l_t)\,,
\cr
Q^{(3)}_{Hl}&=(H^\dag\oLR{D}^I \gamma_\mu H)(\bar{l}_p\tau^I\gamma^\mu l_r)\,.
}
In our case, only the first is relevant and we obtain for the deviation to $G_\mu=G_F^{\rm SM}+\delta G_F$,
\eq{
\label{delta.GF}
\frac{\delta G_F}{G_F^{\rm SM}}
=\frac{1}{\sqrt{2}G_F^{\rm SM}}\frac{|g_{e\mu}|^2}{\tM_2^2}\,,
}
where
\eq{
\frac{1}{\tM_2^2}\equiv \frac{s_\gamma^2}{M_{S_1}^2}+\frac{c_\gamma^2}{M_{S_2}^2}\,.
}
Note that the deviation \eqref{delta.GF} is positive definite.
As the coupling $g_{\alpha\beta}$ modifies the rates to all decays $\ell_\alpha\to \ell_\beta\nu\nu$,
it is convenient to define the deviations\,\cite{crivellin:singly.charged}
\eq{
\label{delta:ab}
\delta_{\alpha\beta}\equiv 
\frac{1}{\sqrt{2}G_F^{\rm SM}}\frac{|g_{\alpha\beta}|^2}{\tM_2^2}
\,,
}
and the relative deviation to $\delta G_F$ \eqref{delta.GF} is simply $\delta_{e\mu}$.
So further constraints arise from lepton universality which will be discussed in the next section.

In contrast to the leptons, the Fermi constant to the quarks $G_\beta=G_{\mu}^{\rm SM}$ is not modified and comparison to the measured 
$G_\mu$ relates
\eq{
|V_{ud}|^2 + |V_{us}|^2 + |V_{ub}|^2 = \left(\frac{G_F^{\rm SM}}{G_\mu}\right)^2
\approx 1-2\delta_{e\mu}\,.
}
The combined value $|V_{ud}|^2 + |V_{us}|^2 + |V_{ub}|^2 = 0.9984 \pm 0.0007$\,\cite{pdg} currently 
shows a deficit which is in tension with the unitarity of the CKM first row\,\cite{cabibbo.ano,ckm.unit}.
It was shown in Ref.\,\cite{crivellin:singly.charged} that this deficit can be explained by the coupling 
$g_{\alpha\beta}$ with the charged scalars where a combined fit took into account other electroweak constraints,
different extractions of the CKM elements and lepton universality tests.
We discuss this fit together with lepton universality constraints in the next subsection.

\subsection{Lepton universality}
\label{sec:lepton.univ}

The couplings of the charged scalars to leptons contribute to the decays $\ell\to 
\ell'\nu\nu$ inducing deviations to lepton flavor universality (LFU) in the SM due to the 
universal coupling of leptons to $W$.
This constraint can be tested in terms of ratios of the ``effective Fermi constants'' 
$G_{\alpha\beta}$ for the different channels\,\cite{nebot,pich:tau}:
\eqali{
\left(\frac{\Gamma_{\tau\to \mu}}{\Gamma_{\tau\to e}}\right)^{\sfrac{1}{4}}
&\propto
\frac{G_{\tau\mu}}{G_{\tau e}}
\approx 1+\frac{1}{\sqrt{2}G_F}\frac{|g_{\mu\tau}|^2-|g_{e\tau}|^2}{\tM_2^2}
=1+\delta_{\mu\tau}-\delta_{e\tau}
\,,\cr
\left(\frac{\Gamma_{\tau\to \mu}}{\Gamma_{\mu\to e}}\right)^{\sfrac{1}{4}}
&\propto
\frac{G_{\tau\mu}}{G_{\mu e}}
\approx 1+\frac{1}{\sqrt{2}G_F}\frac{|g_{\mu\tau}|^2-|g_{e\mu}|^2}{\tM_2^2}
=1+\delta_{\mu\tau}-\delta_{e\mu}
\,,\cr
\left(\frac{\Gamma_{\tau\to e}}{\Gamma_{\mu\to e}}\right)^{\sfrac{1}{4}}
&\propto
\frac{G_{\tau e}}{G_{\mu e}}
\approx 1+\frac{1}{\sqrt{2}G_F}\frac{|g_{e\tau}|^2-|g_{e\mu}|^2}{\tM_2^2}
=1+\delta_{e\tau}-\delta_{e\mu}
\,.
}
Currently, the best-fit values for these quantities are\,\cite{pich:tau:2} $1.0018\pm 0.0032$, 
$1.0030\pm 0.0030$, $1.0011\pm 0.0030$, respectively, and they show hints of LFU violation (LFUV) in 
the $\tau\to \mu$ channel.
We take the errors at 2$\sigma$.

In Ref.\,\cite{crivellin:singly.charged}, it was shown that the CAA and LFUV above can be solved 
with the presence of the singly charged scalar with the coupling $g_{\alpha\beta}$.
The solution is presented there in terms of the deviations \eqref{delta:ab}
where we adapt the scalar mass to the correct combination of our charged scalars.
Following the solution of Ref.\,\cite{crivellin:singly.charged}, 
we will assume that these anomalies are solved if
\eq{
\label{crivellin.fit}
\delta_{e\mu}\in[0.0005,0.0008]\,,\quad
\delta_{\mu\tau}\in[0.0012,0.0042]\,,
}
where we take a rectangular region for simplicity.
The same fit also led to $\delta_{e\tau}\approx 0$ which we assume and is compatible with 
our neutrino mass matrix \eqref{Mnu:ss+gn+zee} with $e=0$.
The constraint \eqref{crivellin.fit} then leads to
\eqali{
\label{cabibbo:gmutau}
0.036\times \frac{\tM_2}{400\unit{GeV}} &<|g_{e\mu}|< 
0.046\times\frac{\tM_2}{400\unit{GeV}}
\,,
\cr
0.056\times \frac{\tM_2}{400\unit{GeV}} &<|g_{\mu\tau}|< 
0.11\times\frac{\tM_2}{400\unit{GeV}}\,.
}
These values for $g_{e\mu},g_{\mu\tau}$ are much larger than the typical couplings in the Zee model\,\cite{herrero}.
Given the rectangular region \eqref{crivellin.fit}, it is also clear that $\delta_{e\mu}/\delta_{\mu\tau}$
are limited to a range. This range can then be translated through \eqref{h/f:g} to a range in $|h|/|f|$
which was shown in \eqref{h/f:range}.

Now, given the range for $\delta_{\mu\tau}$ in \eqref{crivellin.fit} and the range for $|f|$ in 
\eqref{range:m1.f} in our non-aligned model, cf.~Sec.\,\ref{sec:non-aligned}, we can obtain constraints on the 
overall parameter that enter $f$ in \eqref{efh}.
Rewriting $|f|$ as
\eq{
|f|=\frac{A_{\rm Zee}\zeta_e}{v^2}m_\tau^2 \tM_2 \sqrt{\delta_{\mu\tau}}\,,
}
we obtain
\eq{
A_{\rm Zee}\zeta_e\frac{\tM_2}{400\,\unit{GeV}}
\in [3.3,7.5]\times 10^{-8}\,.
}
where $A_{\rm Zee}$ was defined in \eqref{def:AZee}.
So for $\tM_2\sim 400\,\unit{GeV}$, $A_{\rm Zee}\zeta_e\sim 10^{-7}\text{--}10^{-8}$. 
We typically need large $t_\beta$ or small $\zeta_e$ because $s_{2\gamma}$ cannot be much suppressed in order to solve 
$a_\mu$.
As discussed in \eqref{AZee:typical}, these values are necessary for the Zee mechanism if $g_{\alpha\beta}$ is not much 
suppressed.

We also note that lepton flavor universality tests will be improved in the future in the proposed experiment PIONEER 
\cite{PIONEER:2025idw}
which aims to increase the precision in the measurement of $R_{e/\mu}=\Gamma(\pi^+\to e^+\nu(\gamma))/\Gamma(\pi^+\to 
\mu^+\nu(\gamma))$ by a factor of 15.

\subsection{$W$ boson mass}
\label{sec:W.mass}

Considering the usual SM input parameters $\{G_\mu,\alpha,m_Z\}$, we have seen that the Fermi constant $G_\mu$ measured
from muon decay deviates from the SM value by the contribution of charged scalars at
\emph{tree-level} in \eqref{delta.GF}.
New physics also affect the electroweak precision observables through modifications at \emph{one-loop}
to the gauge boson self-energies which can be quantified by the oblique parameters\,\cite{peskin.takeuchi} $S,T$, where 
we neglect $U$.
The deviation to the (pole) $W$ boson mass can thus be written in terms of $S,T$\,\cite{burgess} with the 
additional contribution coming from the tree-level modification $\delta G_F$ in \eqref{delta.GF}.
These contributions lead to the deviation
\eq{
\frac{\delta M^2_W}{M^{2}_W}=
-\frac{\alpha S}{2(c_w^2-s_w^2)}+\frac{c_w^2\alpha T}{c_w^2-s_w^2}
-\frac{s_w^2}{c_w^2-s_w^2}\frac{\delta G_F}{G_F}\,.
}
The last contribution coincides with calculations in similar models such as the model with only one singly charged scalar coupled to the lepton doublets $\ell$\,\cite{herrero:singly.charged}
and also with the muon specific 2HDM\,\cite{muon.specific}.
This contribution was not included in the analysis of the Zee model in Ref.\,\cite{heeck:zee}.
Being negative definite, it also worsens the compatibility with the CDF measurement of the $W$ boson 
mass\,\cite{CDF:2022hxs}.
However, differently from the case with only one singly charged scalar\,\cite{herrero:singly.charged,bilenky}, 
the presence of the second Higgs doublet in our case allows the deviation in $S,T$ to compensate the 
negative definite contribution of $\delta G_F$.
We make use of the expression for $S,T$ from Ref.\,\cite{herrero} which involves the exchange of gauge bosons,  
neutral scalars $h^0,H^0,A^0$ and charged scalars $S_1^+,S_2^+$. 

Currently, the prediction of the SM for the $W$ boson mass is
\eq{
M_W=80.356 \pm 0.005\,,
}
which is compatible with the current global fit\,\cite{pdg} 
\eq{
M_W=80.3692\pm 0.0133\,.
}
In contrast, the value reported by CDF\,\cite{CDF:2022hxs},
\eq{
M_W=
80.4335\pm 0.0094,
}
deviates from the SM prediction by many standard deviations.
In relative terms, the two deviations with respect to the SM predictions are
\eq{
\label{W.mass:dev}
\frac{\delta M_W}{M_W}\Big|_{\rm global}=
(1.6\pm 1.8)\times 10^{-4}\,,
\quad
\frac{\delta M_W}{M_W}\Big|_{\rm CDF}=(0.96\pm 0.13)\times 10^{-3}\,.
}
We will consider these two scenarios for the $W$ mass.

Currently, the electroweak precision observables constrain the oblique parameters $S,T$ to\,\cite{pdg}:
\eq{
\label{ST:pdg}
S\in [-0.12,0.02]\,,\quad
T\in [-0.06,0.06]\,,
}
with correlation of 0.93. We are assuming $U=0$.

\subsection{Neutrino non-standard interactions}

The first operator in \eqref{smeft.op} that induces $\delta G_F$ leads at low energy to the neutral current operator
\eq{
\lag^{\rm NSI}_{\rm d=6}=-2\sqrt{2}\GFSM\eps^{\alpha'\beta'}_{\alpha\beta}
(\bar{\nu}_{\alpha L}\gamma^\mu\nu_{\beta L})
(\bar{e}_{\alpha' L}\gamma_\mu e_{\beta' L})
\,,
}
where the coefficients
\eq{
\eps^{\alpha'\beta'}_{\alpha\beta}\equiv 
-\frac{1}{2\sqrt{2}\GFSM}\big(C_{\underset{\alpha\beta\alpha'\beta'}{ll}}+C_{\underset{\alpha'\beta'\alpha\beta}{ll}}\big)
=
-\frac{1}{2\sqrt{2}\GFSM}\frac{g_{\alpha\alpha'}^*g_{\beta\beta'}}{\tM_2^2}
}
are known as non-standard interaction (NSI) parameters.
The parameters $\delta_{\alpha\beta}$ in \eqref{delta:ab} are then related to them by
\eq{
\delta_{\alpha\beta}=-\eps^{\alpha\beta}_{\alpha\beta}\,.
}
These parameters are constrained by neutrino oscillation data and currently\,\cite{farzan.tortola}
$|\eps^{ee}_{\alpha\beta}|\lesssim 10^{-2}$ at best for propagation in matter where only 
$(\alpha',\beta')=(e,e)$ is relevant.
In our case where $g_{e\tau}=0$, there are no constraints on the mixed term $|g_{e\mu}g_{e\tau}|$
and the constraint \eqref{crivellin.fit} for $(\alpha',\beta')=(\alpha,\beta)$ to solve the Cabibbo angle anomaly
is stronger than NSI constraints.
The sensitivity to the combination
\eq{
\eps^{e\mu}_{\mu\tau}=-\frac{g_{e\mu}^*g_{\mu\tau}}{\sqrt{2}\GFSM\tM_2^2}\,,
}
may reach $7\times 10^{-4}$ using a $2\,\unit{kt}$ OPERA-like near tau detector\,\cite{nsi:tau}.

\section{Results}
\label{sec:results}

\subsection{Aligned case}
\label{sec:results:aligned}

We first discuss the $\Lambda$-aligned case, cf.~\eqref{scenarios}, proposed in Sec.\,\ref{sec:aligned}.
The contribution to the neutrino mass matrix from the tree-level seesaw and GN is aligned, and if we
add the Zee contribution on top, with $g_{e\tau}=0$, we obtain the mass matrix \eqref{Mnu:ss.zee:yuk.a}
with two texture-zeros.
Then the neutrino mass matrix is mostly determined from current neutrino data as in \eqref{aligned:A.B}  and, in particular, the ratio of the Yukawas $\La{2}_\mu$ and $\La{2}_\tau$ are fixed as
\eq{
\frac{|\La{2}_\tau|^2}{|\La{2}_\mu|^2}=\frac{B}{A}\approx 0.8\,.
}
Therefore, assuming the chiral enhanced contribution \eqref{cR:lambda.f} dominates with only 
$f_\mu,\La{2}_\mu,\La{2}_\tau$ nonzero among $f_\alpha$ or $\La{2}_\alpha$,
the branching ratio for the CLFV process $\tau\to \mu\gamma$ is directly related to the 
new physics contribution to $(g-2)_\mu$ as
\eq{
\label{aligned:tau.mu.gamma}
\Br(\tau\to \mu\gamma)=\frac{B}{A}\frac{e^2m_\tau^3}{64\pi\Gamma_\tau m_\mu^2}[\delta a_\mu]^2
\sim 10^{11}\times\frac{B}{A}[\delta a_\mu]^2\,.
}
If we impose the central value \eqref{eq:amuBSM} for $\delta a_\mu$, we get
\eq{
\Br(\tau\to \mu\gamma)\approx 5\times 10^{-7}\,,
}
which violates the current bound in Table \ref{tab:LFV}.
The situation is unchanged even if we allow solving $\delta a_\mu$ within $3\sigma$.
The relation in \eqref{aligned:tau.mu.gamma} assumed that $\La{2}_\mu f_\mu$ is real and positive.
If we allow an imaginary part in $\La{2}_\mu f_\mu$, the branching ratio for $\tau\to\mu\gamma$ 
would be even larger.

So in this $\Lambda$-aligned case with $g_{e\tau}=0$, generating the correct neutrino parameters and solving the $a_\mu$ 
anomaly is not compatible with CLFV constraints.
In contrast, the case with vanishing $b$ but without vanishing $g_{\alpha\beta}$ leads to a two-zero texture
in the entries $(ee)$ and $(\tau\tau)$ that was already incompatible with neutrino parameters\,\cite{2-texture}.
So the property $\La{2}_\mu\sim \La{2}_\tau$ cannot be avoided in other possibilities and this case is 
excluded.

\subsection{Non-aligned case}
\label{sec:results:non}

Here we discuss the final model presented in Sec.\,\ref{sec:non-aligned} where $\La{1},\La{2}$ are 
non-aligned, the latter with muonphilic structure $\La{2}_\alpha=(0,\La{2}_\mu,0)$, while $f_\alpha=(0,f_\mu,0)$ and $g_{e\tau}=0$
for the couplings in \eqref{lag:varphi}.

As benchmark points we consider
\eqali{
\label{bm1}
\text{(BM1)}&\quad 
s_{2\gamma}=0.1\,,~
M_{S_1}=300\,\unit{GeV}\,,~
M_{S_2}=400\,\unit{GeV}\,,~
M_N=10\,\unit{GeV}\,,~
M_{H}\approx M_{A}= 358\,\unit{GeV}\,,
\cr
\text{(BM2)}&\quad 
s_{2\gamma}=0.1\,,~
M_{S_1}=300\,\unit{GeV}\,,~
M_{S_2}=400\,\unit{GeV}\,,~
M_N=10\,\unit{GeV}\,,~
M_{H}\approx M_{A}= 335\,\unit{GeV}\,,
}
where only the masses of the neutral scalars differ.

\begin{figure}[h]
\includegraphics[scale=.8]{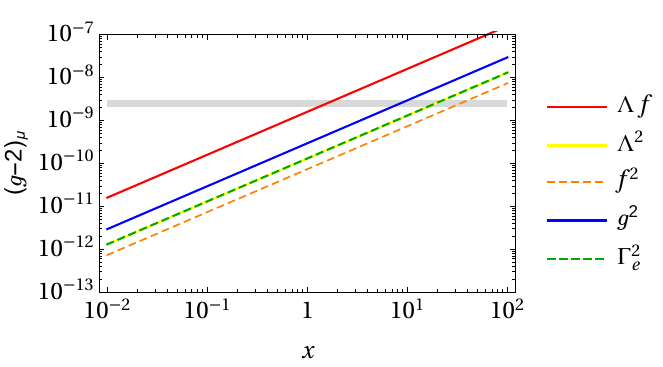}
\caption{\label{fig:g-2.mu}
Separate contributions for $\delta a_\mu=\delta(g-2)_\mu$ from charged scalars in the loop as a function of
$x=|{\La{2}_\mu}^* f_\mu^*|$ (red) in \eqref{cR:lambda.f},
$x=\sum_i|g_{i\mu}|^2$ (blue) in \eqref{cR:g.g},
$x=\sum_i|(\Gamma_e)_{i\mu}|^2$ (yellow) in \eqref{cR:Gamma.Gamma},
$x=|\La{2}_\mu|^2$ (green) in \eqref{cR:lambda.lambda},
$x=|f_\mu|^2$ (orange dashed) in \eqref{cR:f.f}.
The lines in yellow, orange, blue and green show $-\delta a_\mu$ as all the contributions \eqref{cR:lambda.lambda}-\eqref{cR:Gamma.Gamma} have a definite sign and contribute negatively to $\delta a_\mu$.
The gray band denotes the $1\sigma$ band for $\delta a_\mu$ in \eqref{eq:amuBSM}.
The rest of the parameters follows \eqref{bm1}.
}
\end{figure}

Let us start with $a_\mu$.
We show in Fig.\,\ref{fig:g-2.mu} the different contributions \eqref{C.dipole} to $a_\mu$ using the values
in \eqref{bm1} for the masses of $N,S_1,S_2$ and for the mixing angle $\gamma$.
Here we denote each of the contributions as $\Lambda\cdot f,\Lambda^2,f^2,g^2,\Gamma_e^2$ contributions,
respectively.
We can clearly see that the chiral enhanced contribution (red) dominates and an order one $|\La{2}_\mu f_\mu|$ can account for the $a_\mu$ discrepancy while the other contributions are subleading.
Furthermore, with only $\La{2}_\mu,f_\mu\neq 0$, the contributions with $\Lambda\cdot f,\Lambda^2,f^2$ are 
present for $a_\mu$ but not for CLFV.
The coupling $\Gamma_e$ is diagonal and does not contribute to CLFV as well.
So only the contribution $g^2$ will be relevant for CLFV constraints.

\begin{figure}[h]
\includegraphics[scale=.7]{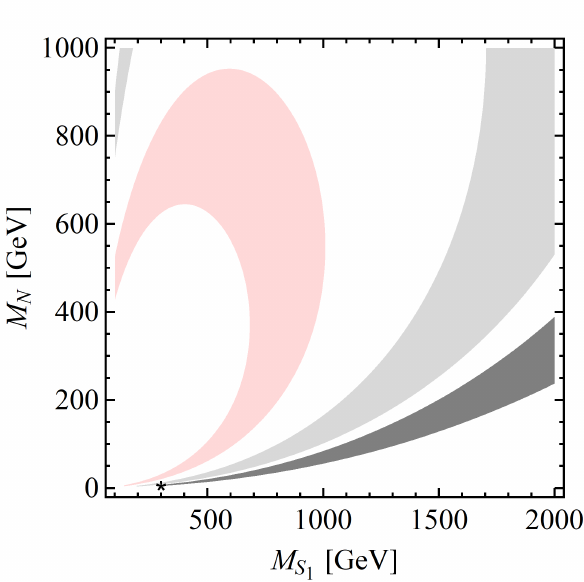}
\caption{\label{fig:g-2.mu:N-S1}
Regions in the $M_N$-$M_{S_1}$ plane solving the $\delta a_\mu$ anomaly \eqref{eq:amuBSM} within $1\sigma$
solely with the chiral enhanced contribution \eqref{cR:lambda.f}
for $M_{S_1}/M_{S_2}=0.9, 3/4, 1/2$ in light red, light gray and gray, respectively.
We choose $s_{2\gamma}=0.1$ and ${\La{2}_\mu}^* f_\mu=1.5$.
The benchmark point \eqref{bm1} is marked with an asterisk.
}
\end{figure}

The chiral enhanced contribution is sufficient to solve $a_\mu$ for a wide range of masses of $N$ and charged scalars
$S_1^+,S_2^+$ away from our benchmark values in \eqref{bm1}.
To illustrate that, we show in Fig.\,\ref{fig:g-2.mu:N-S1} regions in the plane $M_N$-$M_{S_1}$ where $a_\mu$ can be
solved for different values of the ratio $M_{S_1}/M_{S_2}$. It is clear that TeV scale masses for $N$ and $S_i^+$ are
generically possible.
Our benchmark value \eqref{bm1} is also shown for comparison.

\begin{figure}[h]
\includegraphics[scale=1.]{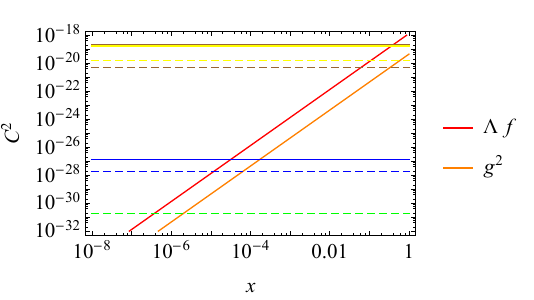}
\caption{\label{fig:clfv}
Separate contributions for $C^2=|C^{\sigma R}_{\beta\alpha}|^2$ ($\unit{GeV^{-2}}$) from 
charged scalars in the 
loop as a function of 
$x=|{\La{2}_\beta}^* f_\alpha^*|$ (red) in \eqref{cR:lambda.f} and
$x=|\sum_ig_{i\beta}^*g_{i\alpha}|$ (orange) in \eqref{cR:g.g}.
The horizontal lines refer to CLFV limits:
current (future) limit on $\Br(\mu\to e\gamma)$ in continuous (dashed) blue; 
future limit from $\mu e$ conversion in Ti nucleus in dashed green;
current (future) limit on $\Br(\tau\to e\gamma)$ in continuous (dashed) yellow; 
current (future) limit on $\Br(\tau\to \mu\gamma)$ in continuous (dashed) brown.
The rest of the parameters follows \eqref{bm1}.
}
\end{figure}

To analyze CLFV, let us consider Fig.\,\ref{fig:clfv} which shows the contributions in \eqref{cR:lambda.f} (red) and
in \eqref{cR:g.g} (orange) for the flavor violating effective coupling in the dipole opeartor \eqref{L.eff:photon}.
For the moment, we allow nonzero values for all couplings $\La{2}_\alpha$ and $f_\alpha$.
We also show the various current and future limits for CLFV radiative decays $\ell\to\ell'\gamma$ in the horizontal
lines.
From the chiral enhanced contribution in red and the current limit in blue of $\mu\to e\gamma$, we can extract that
\eq{
\label{Lambda*f:emu}
\sqrt{|\La{2}_ef_\mu|^2+|\La{2}_\mu f_e|^2}\lesssim 2\times 10^{-5}\,,
}
for the masses in \eqref{bm1}.
It is clear that for order one $f_\mu$ and $\La{2}_\mu$, the couplings $\La{2}_e$ or $f_e$ need to be
at most $10^{-5}$.
The future $\mu e$ conversion limit for the Ti nucleus is expected to improve this limit by two orders
of magnitude.
From the current limit of $\tau\to \mu\gamma$ (brown) we obtain a similar
constraint
\eq{
\label{Lambda*f:mutau}
\sqrt{|\La{2}_\tau f_\mu|^2+|\La{2}_\mu f_\tau|^2}\lesssim 0.4\,.
}
It is clear that the muonphilic structure $\La{2}_\alpha=(0,\La{2}_\mu,0)$ and $f_\alpha=(0,f_\mu,0)$
easily satisfies \eqref{Lambda*f:emu} and \eqref{Lambda*f:mutau}.

\begin{figure}[h]
\includegraphics[scale=.8]{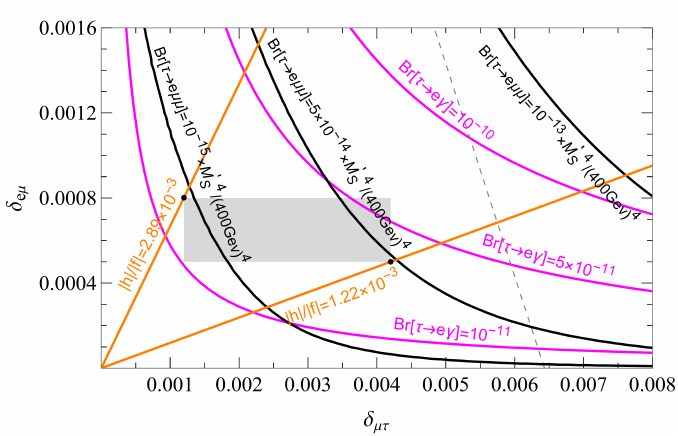}
\caption{\label{fig:deltaemu-mutau}
Preferred region (shaded gray) in the $\delta_{e\mu}$-$\delta_{e\mu}$ plane to solve the Cabibbo anomaly
together with the predictions for $\tau\to e\gamma$
(magenta) and $\tau\to e\mu\mu$ (black).
$M'_S$ is defined in \eqref{def:M'S}.
The orange lines correspond to the limits of \eqref{h/f:range} while the dashed line corresponds to
\eqref{cR:g.g} contributing to $a_\mu$ the value $-5\times 10^{-12}$, i.e., 1\% of the error in \eqref{eq:amuBSM}.
}
\end{figure}

With the muonphilic structure for $\La{2}_\alpha$ and $f_\alpha$, there are no chiral enhanced contributions to flavor
changing processes.
The only remaining flavor changing coupling is
$g_{\alpha\beta}$ which is in fact purely flavor changing.
To solve the Cabibbo anomaly while still passing lepton universality constraints discussed in
Sec.\,\ref{sec:contraints},
we further take $g_{e\tau}=0$, while the pair $g_{e\mu},g_{\mu\tau}$ needs to satisfy
\eqref{cabibbo:gmutau}.
In terms of $\delta_{e\mu},\delta_{\mu\tau}$ in \eqref{delta:ab}, the region is given in \eqref{crivellin.fit} which we
show as a gray rectangle in Fig.\,\ref{fig:deltaemu-mutau}.
In the same figure, we show the branching ratios coming from $\tau\to e\gamma$ and $\tau\to e\mu\mu$. We see that we
are far from the current and even the future limits in Table\,\ref{tab:LFV} and we easily pass these limits.
The lines for $\tau\to e\mu\mu$ depend on the ratio $M'_S$ defined in \eqref{def:M'S} but they become relevant only for
very large masses.
The contribution of $g_{\alpha\beta}$ to $a_\mu$ is negligible and we show in dashed line the values
corresponding to $\delta a_\mu=-5\times 10^{-12}$, i.e., 1\% of the error in \eqref{eq:amuBSM}.
The other CLFV processes, $\mu\to e\gamma$,  $\mu e$ conversion in nuclei, and $\tau\to \mu\gamma$, are absent.

\begin{figure}[h]
\includegraphics[scale=.65]{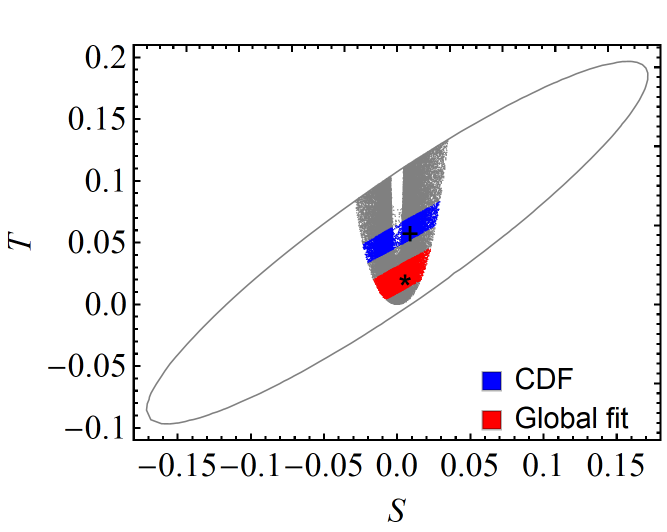}
\caption{\label{fig:S-T}
Scatter plot for the model in the $(S,T)$ plane. The blue points satisfy the deviation in the $W$ mass reported by CDF
while the red points satisfy the current global fit. The values are given in \eqref{W.mass:dev}.
The gray ellipse depict the current global fit within 95\% CL.
}
\end{figure}

As discussed in Sec.\,\ref{sec:W.mass}, we take into account the electroweak precision observables through the oblique 
parameters $S,T$.
Current global fit dictates that the pair $(S,T)$ should be confined within the ellipse \eqref{ST:pdg} at 95\% CL.
This ellipse is depicted in Fig.\,\ref{fig:S-T} in gray.
We generate points varying
\eq{
M_H, M_{S_1}, M_{S_2} \in [100\,\unit{GeV},1\,\unit{TeV}]\,,
\quad
\gamma\in [10^{-4},0.1]\,,
}
with $m_h=125\,\unit{GeV}$ and $M_A=M_H+\delta$, $\delta=10^{-5}\,\unit{GeV}$.
The points that satisfy the $(S,T)$ bound are shown in gray. 
Among them, the points in blue satisfy the $W$ mass deviation reported by CDF while the points in red 
satisfy the current global fit; both of them can be seen in \eqref{W.mass:dev}.
We also show the benchmark points in \eqref{bm1} which satisfy each case.

\begin{figure}[h]
\includegraphics[scale=.65]{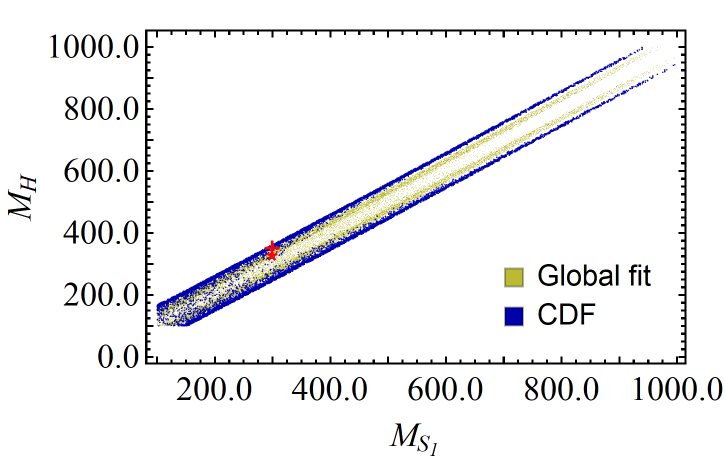}
\caption{\label{fig:MS1.MH}
Masses of $H^0$ and $S_1^+$ satisfying the $S,T$ bounds in Fig.\,\ref{fig:S-T} and  simultaneously satisfying 
the $W$ mass deviation reported by CDF (blue) or satisfying the current global fit (yellow).
}
\end{figure}

The most evident consequence of satisfying these precision electroweak bounds can be seen in Fig.\,\ref{fig:MS1.MH} 
where we only show the points that either satisfy the $W$ mass deviation reported by CDF (blue) or that 
satisfy the current global fit (yellow) corresponding respectively to the blue and red points in Fig.\,\ref{fig:S-T}.
Primarily due to the constraint on $T$ the mass splitting between $H^0$ and $S_1^+$ cannot be so large.
We also show the benchmark points in \eqref{bm1} with the same symbols as in Fig.\,\ref{fig:S-T}.
A similar model that solves the $W$ mass anomaly with dark matter candidates can be seen in Ref.\,\cite{Dcruz:2022dao}.

Let us now briefly discuss the collider constraints and signatures of our model. 
For small charged scalar masses, collider constraints apply.
As the singly charged scalar has the same gauge quantum numbers as the right-handed slepton in supersymmetry,
bound from direct searches of smuons and sleptons can be recast and yields\,\cite{crivellin:singly.charged},
\eq{
\label{MSi:limit}
M_{S_i}\gtrsim 200\,\unit{GeV}\,.
}
This stronger than the limits on charged higgses in the 2HDM from LEP\,\cite{ALEPH:2013htx}.
For the other bounds, we mainly focus on hadronic colliders, such as the LHC. In our setting, the coupling between 
quarks and the extra Higgs doublet $\mathcal{H}_2$ is generic, allowing us to suppress the production from pp collisions 
of the CP-odd neutral scalar $A_{0}$, as well as of the charged scalar $H^{+}$. Since we work on the alignment limit, 
the production of the heavy CP-even neutral $H$ is also suppressed. It should be noticed that the charged mass 
eigenstate $S_{i}^{+}$ contains a admixture of $H^{+}$ and $\varphi^{+}$. However, since $\varphi^{+}$ does not couple 
to quarks, the above reasoning still holds. The RHN does not couple to quarks either. In fact, being a singlet under 
$SU(2)$ it does not couple to any of the gauge bosons, except by tiny admixtures with the active neutrinos. In this 
scenario, the RHN can only be produced from decays of the charged scalars.

We consider charged scalars with mass obeying \eqref{MSi:limit},
so they can be produced by pair production through an 
off-shell photon or $Z$ boson. Since in our scenario the $S_{i}^{+}$ mass is significantly larger than the $N_{R}$ 
mass, we expect the decays $S_{i}^{+}\rightarrow N_{R}\, l_{k}$ to occur. They come from two different sources, related 
to the couplings $\Lambda^{(2)}_{i}$, and $f_{i}$. The decay rates are estimated as 
\begin{align}
\Gamma(S_{i}^{-}\rightarrow N_{R}\, e_{jR})\sim \frac{|f_{j}|^{2}}{16\pi}M_{S_{i}}\,,\quad \quad \Gamma(S_{i}^{-}\rightarrow N_{R} e_{jL})\sim \frac{|\Lambda_{j}^{(2)}|^{2}}{16\pi}M_{S_{i}}\,,
\end{align} 
where we omit the mixing between charged fields. We consider a muonphilic scenario, so only decays to muons are
possible. 
The RHN will be a long-lived state due to the tiny admixture with active neutrinos (the estimate from 
\eqref{eq:y_seesaw} leads to $U_{\mu N}\sim 4 \times 10^{-6}$), which could be seen as missing energy (MET)
only.\,\footnote{Current searches at the LHC with displaced vertex
can only constrain mixings $|U_{\mu N}|^{2}\sim 10^{-7}$ or larger~\cite{ATLAS:2025uah}. Searches for same-sign
dileptons processes do not apply
either~\cite{CMS:2018jxx}.}

The charged scalars can decay to active neutrinos directly as well, through the couplings $\Gamma_{e}$, and $g_{ij}$. The decay rates are estimated as  
\begin{align}
\Gamma(S_{i}^{-}\rightarrow \nu_{j}\, e_{jR})\sim \frac{|\Gamma_{e,j}|^{2}}{16\pi}M_{S_{i}}\,,\quad \quad \Gamma(S_{i}^{-}\rightarrow \nu_{k} e_{jL})\sim \frac{|g_{kj}|^{2}}{16\pi}M_{S_{i}}\,.
\end{align} 
Notice that the decay proportional to $\Gamma_{e}$ is diagonal in flavor, while the decay proportional to $g_{ij}$ is 
purely non-diagonal. Nevertheless, for a hadronic collider such as the LHC, this distinction is immaterial. Once again, 
the main signature will be MET.
In the scenarios studied by us, we require $|\La{2}_\mu f_\mu|=O(1)$, while $g_{ij}\sim 
10^{-2}$ and $\left(\Gamma_{e}\right)_{e,\mu,\tau}=(\zeta_e/v) (m_{e},m_{\mu},m_{\tau})$, cf. \eqref{alignment:Gamme}. 
Thus, the charged scalars will predominantly decay to muons and $N_{R}$ and, when $\zeta_e$ is large, also to  
taus and $\nu_{\tau}$.

Concerning the neutral scalars, they can be produced via associated production ($pp \rightarrow S_{i}^{+} H/A$) or pair 
production (from an off-shell $Z$ boson). Their couplings to quarks can be arbitrarily suppressed, while the couplings 
to charged leptons is proportional to $\Gamma_{e}$. 
Thus, the dominant channel is to $\tau\tau$.
Concerning neutral leptons, the decay to them is governed by $\Lambda^{(2)}$ which primarily induce the decays to 
$\nu_\mu\nu_\mu$  or $\nu_\mu N$, the main signature being MET.

\section{Summary}
\label{sec:summary}

Seeking minimal models connecting $(g-2)_\mu$ with neutrino masses, we consider a model with \emph{two} Higgs doublets, 
\emph{a single} charged scalar singlet 
and \emph{a single} RHN $N_R$, i.e., only three additional multiplets compared to the SM.
Neutrino masses are generated from a combination of contributions: 
tree level $N_R$ exchange (seesaw), one-loop exchange of $N_R$ and neutral higgses (Grimus-Neufeld) and one-loop 
exchange of charged leptons and charged higgses (Zee).
In order to explain $(g-2)_\mu$ with a chiral enhanced contribution involving the same mediators and still 
suppressing charged-lepton flavor violating effects, the combination of the three mechanisms are necessary.
We consider the case where the Yukawas of the fermions to the two Higgs doublets are aligned except for the Yukawas of 
the RHN.

The deviation for $(g-2)_\mu$ can be explained in the presence of two Higgs doublets (2HDMs) mainly through  Barr-Zee 
contributions at two-loops involving light neutral higgses
and containing fermion loops with third generation fermions, 
like in the type Y/X 2HDM\cite{2hdm.typeX}, aligned 2HDM\,\cite{Ilisie,Han,Cherchiglia1,Cherchiglia2} or the Zee 
model\,\cite{barman}. The exchange of RHNs plays no role.
Here we consider a different regime, similar to Ref.\,\cite{nu.g-2},
where the one-loop exchange of $N_R$ and charged higgses are crucial.
The difference is that here there is only one RHN instead of two and the Zee contribution to neutrino mass is switched 
on as we allow the coupling $g_{\alpha\beta}$ between the lepton doublets and charged singlet scalar.
In our case, also the Barr-Zee diagrams are negligible because the BSM neutral higgses are not too light.
With many sources to neutrino mass, the coupling $g_{\alpha\beta}$ here is allowed to be larger and in a different 
regime compared to the pure Zee mechanism\,\cite{barman,herrero} which typically requires couplings of the order of 
$10^{-5}\text{--}10^{-7}$.
With relatively large values for $g_{\alpha\beta}$, we are able to explain the Cabibbo angle anomaly and also the $W$ 
mass deviation reported by CDF.

As the flavor changing between $\mu$ and $e$ sectors are switched off, the main CLFV 
processes that are predicted involve 
the decay of tau and the requirement to solve the Cabibbo angle anomaly predicts the rates to a definite range which, 
however, are far from current experimental reach.
The same requirement to solve the Cabibbo angle anomaly also makes the Zee contribution to the neutrino mass 
matrix in the $(e\mu)$ entry to be much smaller than the $(\mu\tau)$ entry.

In summary we explore a minimal extension of the SM with one additional Higgs doublet, one charged scalar singlet and 
one RHN capable of 
simultaneously giving mass to the active neutrinos, solving the $(g-2)_\mu$ anomaly, solving the Cabibbo angle anomaly 
and explaining the $W$ mass deviation reported by CDF.
As the source of lepton flavor violation is brought to low scale, these kind of extensions are subjected to strong 
constraints from CLFV. We have excluded nonviable variants where the Zee contribution was negligible or the case of 
aligned neutrino Yukawas.






\acknowledgments

A.C.\ acknowledges support from National Council for Scientific and Technological Development – CNPq through project 	102263/2024-8.
R.E.A.B\ acknowledges financial support by the Coordenação de Aperfeiçoamento de Pessoal de Nível Superior - Brasil 
(CAPES) - Finance Code 001.
C.C.N.\ acknowledges partial support from CNPq, grant 312866/2022-4.

\appendix
\section{Analytic diagonalization of $2\times 2$ Majorana mass matrix}
\label{ap:2x2}

A Majorana mass matrix $M$ can always be diagonalized (Takagi factorization) by a unitary matrix $U$ as
\eq{
U^\tp M U=\text{diagonal}\,.
}
For a $2\times 2$ complex symmetric matrix
\eq{
\label{2x2:M}
M=\mtrx{a&c\cr c&b}\,,
}
with real $c>0$, the matrix $U$ can be found \emph{analytically} as
\eq{
\label{2x2:U:formula}
U=\mtrx{e^{i\varphi/2}&\cr & e^{-i\varphi/2}}
\mtrx{c_\theta & s_\theta \cr -s_\theta & c_\theta}
\,,
}
where 
\eq{
\label{2x2:formula:phi}
\varphi=\arg(b+a^*)\,,
\quad
2\theta=\arg(r+ic)\,,
\quad
r=\frac{|b|^2-|a|^2}{2|b+a^*|}\,.
}
The diagonalized but still complex matrix is
\eq{
\label{2x2:masses}
U^\tp MU
=\diag(\mu_1,\mu_2)
=\mtrx{x-\sqrt{r^2+c^2}& \cr &x+\sqrt{r^2+c^2}}\,,
}
with complex
\eq{
x=\ums{2}|b+a^*|+i\frac{\im(ab)}{|b+a^*|}\,.
}
The masses in \eqref{2x2:masses} obey $m_1\le m_2$ with $m_i=|\mu_i|$ and degeneracy is avoided only if both $|b+a^*|\neq 0$ and $\sqrt{r^2+c^2}\neq 0$.
Complex $c$ can be taken into account by an initial global rephasing of $M$.
So one needs to multiply $U$ in \eqref{2x2:U:formula} by $e^{-i\arg(c)/2}$, and consider that the entries $a,b,c$ in the formulas above are taken \emph{after} the rephasing.
For massless lightest neutrino, this additional phase is unphysical and can be dropped.

Parametrization may be simpler if we use the following 5 free parameters: $R,R',\varphi,\varphi'$ and $c$.
The new parameters are defined by
\eq{
\label{R.R'}
b+a^*=2Re^{i\varphi}\,,\quad
b-a^*=2R'e^{i(\varphi+\varphi')}\,,
}
where $\varphi$ is the same as in \eqref{2x2:formula:phi}.
Then
\eq{
\label{r.x}
r=R'c_{\varphi'}\,,\quad
x=R+R'is_{\varphi'}\,.
}
The range is the usual $R,R'\in [0,\infty)$ and $\varphi,\varphi'\in [0,2\pi)$.

We now show how to obtain these formulas.
We first determine when $M$ in \eqref{2x2:M} can be diagonalized by a \emph{real} orthogonal matrix, i.e., its real and imaginary parts can be diagonalized simultaneously.
The condition is
\eq{
\big[\re(M),\im(M)\big]=\ums[i]{2}[M,M^*]=\bs{0}_{2\times 2}\,,
}
and it implies
\eq{
\im[(b-a)c^*]=0\,.
}
In the basis where $c$ is real, we need real $a-b$.
This will not be satisfied in general but we can always apply a rephasing
\eq{
M'=\mtrx{e^{i\varphi/2}&\cr & e^{-i\varphi/2}}M\mtrx{e^{i\varphi/2}&\cr & e^{-i\varphi/2}}
=\mtrx{a'&c\cr c&b'}
\,.
}
Imposing real $a'-b'$ determines $\varphi$ as in \eqref{2x2:formula:phi}.
With $\varphi$ determined, the variables $r,x$ are fixed by
\eq{
r=-\ums{2}\re(a'-b')\,,\quad
x=\ums{2}(a'+b')\,,
}
and $M'$ becomes
\eq{
M'=x\id_2+\mtrx{-r&c\cr c&r}\,.
}
The first matrix is invariant by orthogonal transformations and the second matrix, which is real, can be diagonalized by the second matrix in \eqref{2x2:U:formula} with $\theta$ given in \eqref{2x2:formula:phi}.

\section{Masses as input: general case}
\label{ap:masses}

It is often more convenient to use the masses $m_1,m_2$ as input.
We denote the diagonal entries in \eqref{2x2:masses} as the complex masses $\mu_1,\mu_2$. The masses are the moduli $m_i=|\mu_i|$.

We will show here that we can use $(\mu_1,\mu_2,c,\varphi)$ as the five input parameters instead of $a,b,c$
in the basis where $c$ in \eqref{2x2:M} is real.
There are three parameters in the complex masses $\mu_i$ because they obey $\im(\mu_1)=\im(\mu_2)$, cf.\,\eqref{2x2:masses}. So we can write
\eq{
\label{mu12:y}
\mu_1=x_1+iy\,,\quad
\mu_2=x_2+iy\,,
}
where $y$ is not related to the Yukawa coupling in \eqref{mD:y}.
The two parameters $x_1^2,x_2^2$ are fixed by the masses $m_1,m_2$ in
\eq{
|\mu_1|^2=m_1^2\,,\quad
|\mu_2|^2=m_2^2\,,
}
where $x_2>0$ and the sign of $x_1$ is free. The parameter $y$ is also free restricted to $|y|\le m_1$.
The parameter $R$ in \eqref{R.R'} is obtained from
\eq{
R=\frac{x_1+x_2}{2}\,,
}
while $\varphi\in [0,2\pi)$ is free.
These two quantities determine $b+a^*$ in \eqref{R.R'}.
The quantity $r$ in \eqref{r.x} is given by
\eq{
r^2=(x_2-R)^2-c^2\,,
}
where $c>0$ is free, restricted to $|c|\le |x_2-R|$.
The sign of $r$ is also free.
From \eqref{r.x}, we can determine
\eq{
\label{R':y}
R'e^{i\varphi'}=r+iy\,.
}
This quantity together with $\varphi$ of input determine $b-a^*$, cf.\,\eqref{R.R'}.
The original parameters $(a,b,c)$ are then determined.
The diagonalization matrix \eqref{2x2:U:formula} can be also determined from \eqref{2x2:formula:phi}, adding the Majorana phases from the phases of $\mu_1,\mu_2$.

\begin{figure}[h]
\includegraphics[scale=.15]{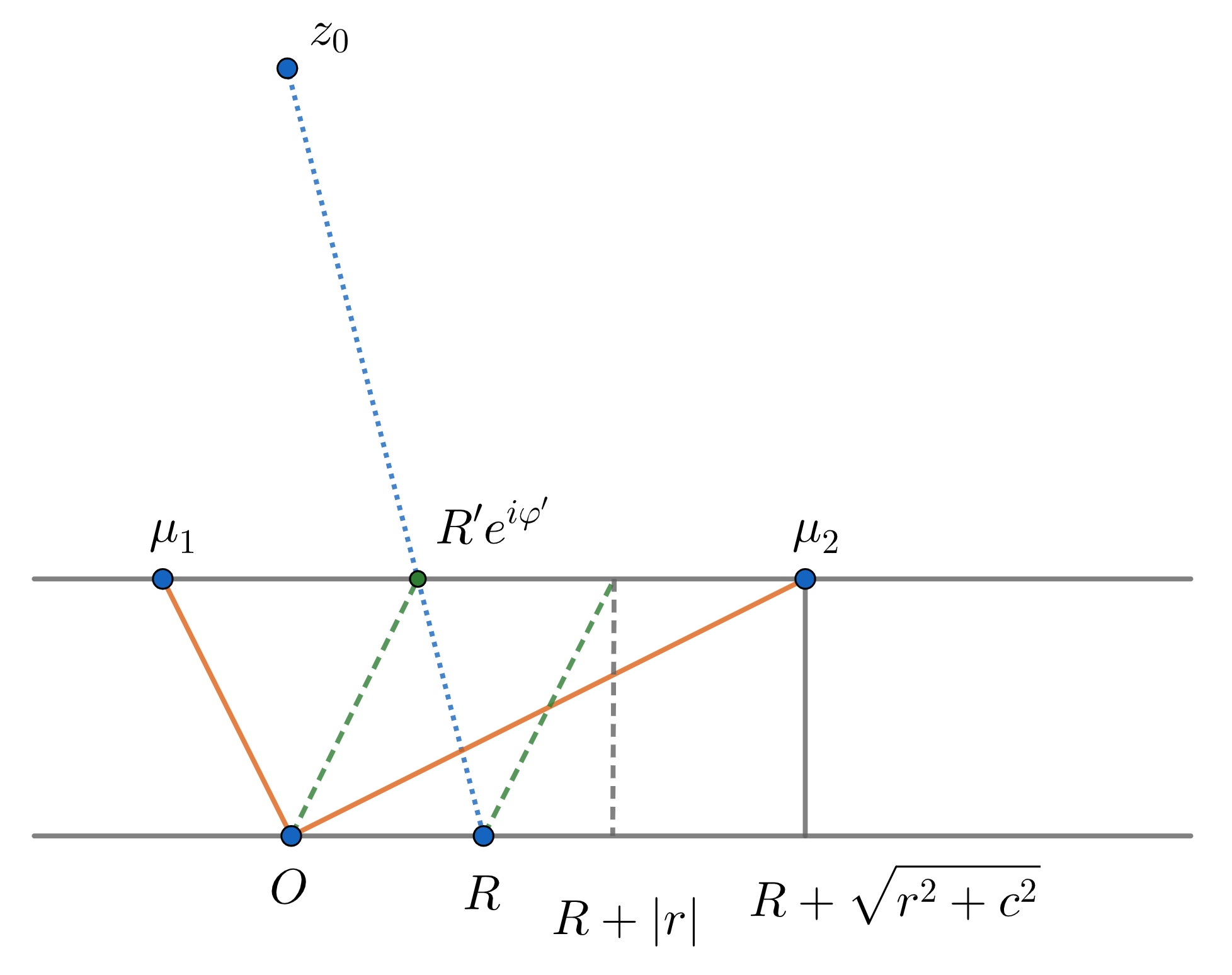}
\caption{\label{fig:mu12}%
Parameters in the complex plane for using the complex masses $\mu_1,\mu_2$ in \eqref{mu12:y} as input.
The derived parameters $R,r,R'e^{i\varphi'}$ are also shown.
The point $z_0$ in \eqref{z0} and the dotted line joining $z_0$-$R$ is only relevant for GN; see appendix \ref{ap:GMN}.
In this figure, $x_1<0$, $y>0$, $r>0$ in \eqref{R':y} and $\kappa>0$ in \eqref{line:kappa}.
}
\end{figure}

The geometrical construction is depicted in Fig.\,\ref{fig:mu12}.
We show the complex masses $\mu_1,\mu_2$, and the points $R,R'e^{i\varphi'}$.

\section{Masses as input: GN mechanism}
\label{ap:GMN}

In the GN mechanism, the effective $2\times 2$ mass matrix \eqref{Sigma:GNM} depends on four parameters
$(m_0,\sqrt{\LGN}d,\sqrt{\LGN}d',\alpha)$ and it is not a generic $2\times 2$ Majorana mass matrix.
Because of that, we cannot directly use the result in appendix \ref{ap:masses}.
We show here that we can use the four parameters $(\mu_1,\mu_2,c)$ as input while $\varphi$ will be a dependent parameter.
The tree-level mass $m_0$ will be also a dependent quantity.

The quantities $R,R'e^{i\varphi'}$ are fixed as in appendix \ref{ap:masses}.
The quantity
\eq{
\label{z0}
z_0\equiv \frac{m_0}{2}e^{-i\varphi}\,,
}
will be determined from the construction in Fig.\,\ref{fig:mu12}.
The dotted line depicts the algebraic relation
\eq{
\label{line:kappa}
R'e^{i\varphi'}=(1-\kappa)z_0+\kappa R\,,
\quad
\kappa\in [-1,1]\,.
}
The quantity $\kappa$ is determined from
\eq{
\label{radii:c}
|R-z_0|^2-|R'e^{i\varphi'}-z_0|^2=c^2\,,
}
which is solved by
\eq{
\kappa=\frac{A-1}{A+1}\,,
\quad A\equiv \frac{c^2}{|R-R'e^{i\varphi'}|^2}\,.
}
The dotted line segment has $R'e^{i\varphi'}$ between the points $R$ and $z_0$ when $\kappa>0$ while $z_0$ lies between $R'e^{i\varphi'}$ and $R$ when $\kappa<0$.

Now we deduce eqs.\,\eqref{line:kappa} and \eqref{radii:c}.
Comparing \eqref{Sigma:GNM} to \eqref{2x2:M}, we identify
\eqali{
b+a^*&= m_0 +\LGN e^{-i\alpha}(d'^2+d^2)=2Re^{i\varphi}\,,
\cr
b-a^*&= m_0 +\LGN e^{-i\alpha}(d'^2-d^2)=2R'e^{i\varphi'}e^{i\varphi}\,,
\cr
c&=|\LGN dd'|\,.
}
Then it is clear that \eqref{radii:c} is satisfied.
It is also clear that
\eqali{
R'e^{i\varphi'}-z_0
=\frac{\LGN}{2} e^{-i(\alpha+\varphi)}(d'^2-d^2)
=\kappa (R-z_0)\,,
}
with
\eq{
\kappa =\frac{d'^2-d^2}{d'^2+d^2}\,.
}
This implies \eqref{line:kappa} with $\kappa\in[-1,1]$.

We can also impose that the one-loop contribution does not exceed 50\% compared to the tree-level mass $m_0$\,\cite{stockinger}. This leads to the condition
\eq{
|\mu_1| < m_0 < 2|\mu_2|\,.
}


\end{document}